\documentclass[amsmath,amssymb,twocolumn]{aastex63}
\usepackage{color}
\usepackage{amsmath}
 \usepackage{ulem}

\begin{document}


\title{A low-mass binary neutron star: long-term ejecta evolution and kilonovae with weak blue emission}


\author{Kyohei Kawaguchi}
\affil{Institute for Cosmic Ray Research, The University of Tokyo, 5-1-5 Kashiwanoha, Kashiwa, Chiba 277-8582, Japan}
\affiliation{Center for Gravitational Physics,
 Yukawa Institute for Theoretical Physics, 
Kyoto University, Kyoto, 606-8502, Japan} 
\author{Sho Fujibayashi}
\affil{Max Planck Institute for Gravitational Physics (Albert Einstein Institute), Am M\"{u}hlenberg 1, Potsdam-Golm, 14476, Germany}
\author{Masaru Shibata}
\affil{Max Planck Institute for Gravitational Physics (Albert Einstein Institute), Am M\"{u}hlenberg 1, Potsdam-Golm, 14476, Germany}
\affiliation{Center for Gravitational Physics,
 Yukawa Institute for Theoretical Physics, 
Kyoto University, Kyoto, 606-8502, Japan} 

\author{Masaomi Tanaka}
\affil{Astronomical Institute, Tohoku University, Aoba, Sendai 980-8578, Japan}
\and
\author{Shinya Wanajo}
\affil{Max Planck Institute for Gravitational Physics (Albert Einstein Institute), Am M\"{u}hlenberg 1, Potsdam-Golm, 14476, Germany}
\affiliation{Interdisciplinary Theoretical and Mathematical Science (iTHEMS) Research Group, RIKEN, Wako, Saitama, 351-0198, Japan}

\newcommand{\angstrom}{\text{\normalfont\AA}}
\newcommand{\rednote}[1]{{\color{red} (#1)}}

\begin{abstract}
We study the long-term evolution of ejecta formed in a binary neutron star (BNS) merger that results in a long-lived remnant NS by performing a hydrodynamics simulation with the outflow data of a numerical relativity simulation as the initial condition. At the homologously expanding phase, the total ejecta mass reaches $\approx0.1\,M_\odot$ with an average velocity of $\approx0.1\,c$ and lanthanide fraction of $\approx 0.005$. We further perform the radiative transfer simulation employing the obtained ejecta profile. We find that, contrary to a naive expectation from the large ejecta mass and low lanthanide fraction, the optical emission is not as bright as that in GW170817/AT2017gfo, while the infrared emission can be brighter. This light curve property is attributed to preferential diffusion of photons toward the equatorial direction due to the prolate ejecta morphology, large opacity contribution of Zr, Y, and lanthanides, and low specific heating rate of the ejecta. Our results suggest that these light curve features could be used as an indicator for the presence of a long-lived remnant NS. We also found that the bright optical emission broadly consistent with GW170817/AT2017gfo is realized for the case that the high-velocity ejecta components in the polar region are suppressed. These results suggest that the remnant in GW170817/AT2017gfo is unlikely to be a long-lived NS, but might have collapsed to a black hole within ${\cal O}(0.1)$ s.
\end{abstract}


\section{Introduction}\label{sec:intro}
Binary neutron star (BNS) mergers are interesting phenomena in various astrophysical aspects. Gravitational waves from BNS are among the most robust targets for the ground-based gravitational-wave detectors~(LIGO:~\citealt{TheLIGOScientific:2014jea}, Virgo:~\citealt{TheVirgo:2014hva}, KAGRA:~\citealt{Kuroda:2010zzb}). A fraction of the NS material can be ejected from the system during the merger~\cite[e.g.,][]{Rosswog:1998hy,Ruffert:2001gf,Hotokezaka:2012ze}, and the so-called $r$-process nucleosynthesis is expected to take place in such ejected material. The ejecta launched from BNS mergers are considered as among the robust production site for the about half of the elements heavier than iron in the universe~\citep{Lattimer:1974slx,Eichler:1989ve,Freiburghaus1999a,Cowan:2019pkx}. 
 BNS mergers are also expected to be associated with various electromagnetic (EM) transients in a wide range of wavelengths~\citep[e.g.,][]{1984SvAL...10..177B,1986ApJ...308L..43P,Eichler:1989ve,Li:1998bw,Metzger:2011bv,Hotokezaka:2015eja,Carrasco:2020sxg}. The simultaneous detection of the EM counterparts to the gravitational wave detection enables us to identify the host galaxy of the merger event. 

Among various possible EM counterparts, a kilonova is one of the most promising EM transients, which occurs as a result of the mass ejection during the merger and the post-merger process~\citep{Li:1998bw,Kulkarni:2005jw,Metzger:2010sy,Kasen:2013xka,Tanaka:2013ana}. Nuclei synthesized in the ejected material include many radioactive isotopes, and hence, the ejected material is heated by the radioactive decay of such isotopes. Thermal photons emitted in the heated material propagate through the expanding ejecta interacting with the synthesized elements, and the emission diffused out from the ejecta is observed as a kilonova. Since the elements synthesized in the NS merger ejecta, particularly the lanthanides, have large contribution to the opacity, the kilonova emission is expected to be bright in the optical and near-infrared wavelengths~\citep{Kasen:2013xka,Tanaka:2013ana,Kasen:2014toa,Wollaeger:2017ahm,Gaigalas:2019ptx,Tanaka:2019iqp}. The optical and infrared EM counterparts associated with the first gravitational wave event from a BNS merger, AT2017gfo,~\citep{Andreoni:2017ppd,Arcavi:2017xiz,Coulter:2017wya,Cowperthwaite:2017dyu,Diaz:2017uch,Drout:2017ijr,
Evans:2017mmy,Hu:2017tlb,Valenti:2017ngx,Kasliwal:2017ngb,Lipunov:2017dwd,Pian:2017gtc,Pozanenko:2017jrn,Smartt:2017fuw,Tanvir:2017pws,Troja:2017nqp,Utsumi:2017cti} was indeed found to be consistent with the prediction for kilonova emission~\citep[e.g.,][]{Cowperthwaite:2017dyu,Kasen:2017sxr,Kasliwal:2017ngb,Perego:2017wtu,Tanaka:2017qxj,Villar:2017wcc,Rosswog:2017sdn,Kawaguchi:2018ptg}, while the origin and the property of the ejecta are still under debate.

The kilonova light curves are characterized by the mass, velocity, radioactive heating, and opacity of ejecta~\citep{Li:1998bw,Kasen:2013xka,Kasen:2014toa,Barnes:2016umi,Wollaeger:2017ahm,Tanaka:2017qxj,Tanaka:2017lxb}. The isotopic abundances which determine the radioactive heating rate and opacity of the ejecta reflect the conditions under which the ejecta are formed, such as the expansion velocity, entropy, and electron fraction ($Y_{\rm e}$)~\citep{Wanajo:2014wha,Wu:2016pnw,Lippuner:2017bfm,Wanajo:2018wra}. The morphology of ejecta, which is also pointed out to be an important factor to characterize the resulting light curves~\citep{Kasen:2014toa,Kawaguchi:2018ptg,Kawaguchi:2019nju,Bulla:2019muo,Zhu:2020inc,Darbha:2020lhz,Korobkin:2020spe}, also reflects the mass ejection history. Thus, the detailed observation of the kilonova emission can be used as a probe for the underlying physics in the merger process. For this purpose, the quantitative understanding of the ejecta property as well as the accurate prediction of the kilonova light curves is crucial.

Many efforts have been made to investigate the ejecta property by performing numerical-relativity (NR) simulations~\cite[see e.g.,][ for a review]{Shibata:2019wef}. The recent NR simulations and nucleosynthesis calculations have revealed the detailed property of the ejected material and the resulting element abundances together with the dependence on the mass ejection mechanism and the binary parameters, such as the NS mass and NS equation of state (EOS)~\citep{Hotokezaka:2012ze,Bauswein:2013yna,Sekiguchi:2015dma,Sekiguchi:2016bjd,Radice:2016dwd,Dietrich:2016hky,Bovard:2017mvn,Kiuchi:2017zzg,Dessart:2008zd,Metzger:2014ila,Perego:2014fma,Just:2014fka,Wu:2016pnw,Siegel:2017nub,Shibata:2017xdx,Lippuner:2017bfm,Fujibayashi:2017puw,Siegel:2017jug,Ruiz:2018wah,Fernandez:2018kax,Christie:2019lim,Perego:2019adq,Fujibayashi:2020qda,Fujibayashi:2020jfr,Fujibayashi:2020dvr,Bernuzzi:2020txg,Ciolfi:2020wfx,Vsevolod:2020pak,Wanajo:2014wha,Wanajo:2018wra}. Based on or motivated by the knowledge of the ejecta profile and the element abundances obtained by those simulations, radiative transfer simulations with the realistic heating rate and/or the detailed opacity calculations are performed to predict the kilonova light curves~\citep[e.g.,][]{Kasen:2013xka,Kasen:2014toa,Barnes:2016umi,Wollaeger:2017ahm,Tanaka:2017lxb,Wu:2018mvg,Kawaguchi:2018ptg,Hotokezaka:2019uwo,Kawaguchi:2019nju,Korobkin:2020spe,Bulla:2020jjr,Zhu:2020eyk,Barnes:2020nfi,Nativi:2020moj}, and the application to the real events is also actively conducted~\citep[e.g.,][]{Kasen:2017sxr,Evans:2017mmy,Tanaka:2017qxj,Kawaguchi:2019nju,Coughlin:2019zqi,Andreoni:2019qgh,Coughlin:2020fwx,Kyutoku:2020xka,Kawaguchi:2020osi,Anand:2020eyg}.

However, the ejecta profiles employed in the previous studies are highly simplified. At the time of the kilonova emission, the ejecta are expected to be in the homologously expanding phase, as the internal energy is lost and becomes much smaller than the kinetic energy due to the adiabatic cooling. On the other hand, in the NR simulations, the ejected material escapes from the computational domain during the evolution before it reaches the homologously expanding phase because the size of the domain and the simulation time are limited by the computational cost. At the time of ejecta evaluation, the ejected material still has non-negligible amount of internal energy compared to its kinetic energy, and the ejecta trajectory can be modified during the expansion due to the thermal pressure~\citep{Kastaun:2014fna}. Furthermore, some fraction of material ejected in the later phase could have higher velocity than the ejecta which have already escaped from the computational domain~\citep[e.g.,][]{Ciolfi:2020wfx,Vsevolod:2020pak,Fujibayashi:2020dvr}. Those fluid elements should interact hydrodynamically in the subsequent evolution. Hence, the ejecta profile at the stage of homologous expansion is not trivial only from the output of the NR simulations. 

\cite{Rosswog:2013kqa} and \cite{Grossman:2013lqa} performed pseudo-Newtonian hydrodynamics simulations for BNS mergers, and studied the long-term evolution of the dynamical ejecta component until it reached the homologously expanding phase. However, the recent NR simulations suggest that the dynamical ejecta component accounts only for a minor fraction of ejected material, and the ejecta are dominated by the post-merger ejecta component (e.g.,~\citealt{Hotokezaka:2012ze,Metzger:2014ila, Fernandez:2018kax,Fujibayashi:2020dvr}), in which the contribution from the internal energy to the ejecta expansion could be more significant~\citep{Kastaun:2014fna}. ~\cite{Fernandez:2014bra} and \cite{Fernandez:2016sbf} performed long-term simulations for black hole-NS mergers to investigate the effect of the interplay between the dynamical and post-merger components. They indeed found that the interaction of the multiple ejecta components can modify the ejecta profile as well as the property of the fall-back material. Thus, to predict more realistic kilonova light curves, it is crucial to follow the hydrodynamics evolution of the multiple ejecta components until the homologously expanding phase.

In this work, we investigate the long-term evolution of ejecta and the kilonova emission from BNS mergers for the cases that the merger remnant NS survives for a long period. For this purpose, we employ the outflow data of the NR simulations obtained in~\cite{Fujibayashi:2020dvr} as the initial conditions. First, we perform a hydrodynamics simulation of ejecta until the system reaches the homologously expanding phase. Then, we perform the radiative transfer simulation employing the ejecta profile in the homologously expanding phase obtained by our hydrodynamics simulation, and we discuss the property of the resulting light curves. This paper is organized as follows: In Section~\ref{sec:hydro}, we describe the setup and the results of the long-term hydrodynamics simulation. In Section~\ref{sec:LC}, we show the results of the radiative transfer simulation employing the ejecta profile obtained by the long-term hydrodynamics simulation. We show that a large amount of ejecta with low lanthanide fraction does not necessarily result in bright optical emission, and that the prolate ejecta morphology, spatial distribution of the Zr, Y, and lanthanides, and low heating rate are the keys for this light curve property. In Section~\ref{sec:D}, we discuss the implication for the future kilonova observation based on our findings, indication of the bright optical emission in a kilonova, and possible non-LTE effect on the late-time light curves. Finally, we summarize this paper in Section~\ref{sec:S}. Throughout this paper, $c$ denotes the speed of light.

\section{Hydrodynamics evolution}\label{sec:hydro}
\subsection{Hydrodynamics Simulation}\label{sec:hydro:setup}
~\cite{Fujibayashi:2020dvr} performed NR simulations of BNS mergers for the cases that the remnant massive NS survives until the end of simulations ($\gtrsim4$--6 s after the merger). The NR simulations were performed with two steps: First, the simulations employing a 3D general relativistic neutrino radiation hydrodynamics code were performed to follow the merger phase. Then, to study the post-merger dynamics, the axisymmetric general relativistic neutrino radiation viscous hydrodynamics simulations were performed employing the hydrodynamics configuration of the 3D simulations as the initial data. By this study, the hydrodynamics profile of the ejecta containing both dynamical and post-merger components was obtained consistently; the former was driven in the first $\sim 50$\,ms after the onset of merger by shock heating in the collision surface and/or gravitational torque of the non-axisymmetric merger remnant, while the latter was launched subsequently from the remnant NS-torus system driven by the viscosity and neutrino irradiation~\cite[see e.g.,][and the references therein for a review]{Shibata:2019wef}.

To determine the ejecta profile in the homologously expanding phase, we solve the hydrodynamics evolution of ejecta by employing the axisymmetric outflow data obtained by NR simulations in~\cite{Fujibayashi:2020dvr} as the inner boundary condition (hereafter, we refer to their result and our present result by the terms ``NR simulation" and ``HD simulation", respectively, to distinguish those two). Specifically, we employ the outflow data of the model DD2-125M in~\cite{Fujibayashi:2020dvr} as representative of their models that result in long-lived remnant NSs. In this model, the merger of an equal-mass binary with each NS mass of $1.25M_\odot$ is considered employing a finite-temperature EOS for nuclear matter referred to as DD2~\citep{Banik:2014qja}, and the viscous parameter of the axisymmetric NR simulation was set to be $\alpha_{\rm vis}=0.04$. Note that the dynamical and post-merger ejecta masses for this fiducial model are $\approx10^{-3}\,M_\odot$ and $\approx10^{-1}\,M_\odot$, respectively. The dynamical ejecta mass and torus mass of the fiducial model are in agreement with the predictions of the fitting models by~\cite{Dietrich:2016fpt,Radice:2018pdn,Kruger:2020gig} within the errors of the fits (see also ~\citealt{Coughlin:2018fis,Nedora:2020qtd}). In the following, we refer to the results obtained by employing this outflow data and the setup described below as the fiducial model.

We note that~\cite{Fujibayashi:2020dvr} found similar outcomes (e.g., the mass, velocity, and element pattern of ejecta) for the models with a different EOS (SFHo, ~\citealt{Steiner:2012rk}) or a more massive (single) NS mass ($1.35\,M_\odot$) to those of DD2-125M. Thus, even for different total mass and EOS, we expect that BNS mergers that accompany the formation of long-lived remnant NSs can result in qualitatively the same property of the ejecta profile and kilonova lightcurves as those for the fiducial case (see Section~\ref{sec:D}).
 
On the other hand, a larger amount of heavy nuclei will be synthesized if much shorter mass ejection time scale is realized, for example, by higher viscosity or magneto-hydrodynamical effects~\citep{Fujibayashi:2020qda,Fujibayashi:2020jfr,Fujibayashi:2020dvr}. Indeed, in~\citealt{Fujibayashi:2020dvr}, the NR simulation employing an effectively larger value of the viscous parameter was performed to investigate the ejecta property in such a situation ($\alpha_{\rm vis}=0.10$, referred to as ``DD2-125M-h"), and found the production of heavy $r$-process nuclei (see also the right panel of Figure~\ref{fig:abuntemp}). In this work, a HD simulation as well as a radiative transfer simulation is also performed for the higher viscosity model to study how the mass ejection time scale of the accretion disk surrounding the remnant NS changes the kilonova lightcurves (see Section~\ref{sec:LC:HVRes} and Appendix~\ref{app:HV} for the results). 
While NR simulations employing a smaller viscous parameter are currently not available and beyond the scope of this paper, the speculation for such cases are also discussed in Section~\ref{sec:D}.

To follow the hydrodynamics evolution of ejecta, we develop a new code for solving the relativistic Euler's equation in the spherical coordinates. The detail of the formulation is summarized in Appendix~\ref{app:form}. In the hydrodynamics simulation code, the effect of fixed-background gravity is taken into account by employing the non-rotating black-hole metric in the isotropic coordinates. We set the initial Arnowitt-Deser-Misner mass~\citep{Arnowitt:1960zzc} of the axisymmetric NR simulation $\approx 2.46 M_\odot$ as the black-hole mass of the metric. \footnote{We note that the relative correction of the remnant NS spin to the metric is negligible because it is an order of $\chi^2 (GM/c^2 r_{\rm in})^2\sim 10^{-12}$ with $\chi$, $M$, and  $r_{\rm in}$ being the dimensionless spin, mass of the remnant NS, and inner radius of the computational domain, respectively~\citep{1975ctf..book.....L}.} Following the NR simulation, the axisymmetry and equatorial plane symmetry are imposed for our HD simulation. For the polar angle $\theta$, 128 grid points are set with uniform grid spacing. For the radial direction, non-uniform grid structure is employed to appropriately resolve the ejecta in the homologously expanding phase. More precisely, the $j$-th grid point is given by

\begin{align}
	{\rm ln}\,r_j={\rm ln}\left(\frac{r_{\rm out}}{r_{\rm in}}\right)\frac{j-1}{J}+{\rm ln}\,r_{\rm in},\,j=1\cdots J+1.\label{eq:grid}
\end{align} 
Here, $r_{\rm in}$ and $r_{\rm out}$ denote the inner and outer radii of the computational domain, respectively, and $J$ denotes the total number of the radial grid points. In this work we set $J=1024$, and $r_{\rm in}$ and $r_{\rm out}$ are set to be $r_{\rm ex}$ and $10^3\,r_{\rm ex}$ (see below). We confirm that the fiducial grid resolution employed in our HD simulations is sufficiently fine by checking that the results do not change for the simulation with 2048 and 256 grids for the radial and latitudinal directions, respectively. We refer to the simulation time of the axisymmetric NR simulation as $t$, of which origin corresponds to $\approx50$ ms after the onset of merger, and the same time origin is employed for the present HD simulations.

The inner boundary of the long-term HD simulations is initially set to be $r_{\rm ex}= 8000\,{\rm km}$, which agrees with the extraction radius in the NR simulation from which the ejecta information was obtained. Note that, by this choice of $r_{\rm ex}$, the dynamical ejecta with the velocity $\lesssim 0.6\,c$ are contained inside the extraction radius of the axisymmetric NR simulation at $t=0$, and the mass of the ejecta already escaped from the extraction radius is negligible ($\sim 10^{-7}\,M_\odot$). In the NR simulation, the rest-mass density, $\rho$, velocity of the fluid, $v^i=u^i/u^t\,(i=x,y,z)$, and pressure, $P$, of the ejecta were recorded at the time when ejecta reached the extraction radius, and they were obtained as functions of the simulation time $t$ and the latitudinal angle $\theta$. The rest-mass density and velocity at $r=r_{\rm in}$ are set by employing these data. We employ the ideal-gas EOS with the adiabatic index of $\Gamma=4/3$ assuming that the total pressure is dominated by the radiation pressure. We confirm in the HD simulations that the domination of the radiation pressure indeed holds up to $t=1$ day. The specific internal energy of the fluid at $r=r_{\rm in}$ is set so that the pressure agrees with that of the outflow data. After the NR simulation data are run out at $t \approx 8.5\,{\rm s}$, the HD simulation is continued by setting the floor-value to the density of the inner boundary. The effect of this truncation is discussed below. 

To follow the long-term evolution of the system, we add the radial grid points to the outside of the originally outer boundary when the high velocity edge of the outflow reaches the outer boundary of our HD simulation. At the same time, the innermost radial grid points are removed to keep the total number of the radial grid points. By this prescription, the material with the radial velocity in the range of $10^{-3} <v/c <1$ is always contained in the computational domain. We note that the total mass of the material lost by removing the innermost radial grid points is $<10^{-3}\,M_\odot$, which is less than $1\%$ of the post-merger ejecta mass located around the inner boundary ($\sim0.1\,M_\odot$), and the re-gridding process has a negligible effect on the evolution.

In the NR simulation, nucleosynthesis calculations were performed by using tracer particles (see~\citealt{Fujibayashi:2020dvr} for the detail), and the element abundances and the radioactive heating rate of the fluid element ejected from the system are determined as functions of ejection time and angle by these calculations. Note that, in the NR simulations, the nucleosynthesis calculations were performed by employing the thermodynamical condition at each location of the particle as far as the particles were in the computational domain. After the particles reached $r=r_{\rm ex}$, the calculation was resumed assuming the free expansion. Hence, the detailed hydrodynamical evolution of the fluid elements, such as the shock heating due to the interaction between different ejecta components, was not taken into account in the nucleosynthesis calculations after the particles passed $r=r_{\rm ex}$. Nevertheless, as we see in Section~\ref{sec:hydro:heat}, we expect that the modification by such effects will be only minor.

To determine the spatial distribution of the radioactive heating rates and the element abundances of the ejecta, the injected time when each fluid element reached $r_{\rm in}$ ($t_{\rm in}$) and angle ($\theta_{\rm in}$) of the material are also considered to be variables of the fluid in addition to usual hydrodynamics variables. These values are evolved by solving the following advection equations of the fluid elements in the conservative form:

\begin{align}
	\partial_t\left(\rho_* Q\right)+\partial_i\left(\rho_* Q v^i\right)=0,
\end{align}
where $Q=t_{\rm in}\,{\rm or}\,\theta_{\rm in}$. For each time step of the HD simulation, the radioactive heating in the ejecta is considered by employing the heating rate obtained by the NR simulation referring to the injected time and angle of the fluid element.

\subsection{Results}\label{sec:hydro:res}
\subsubsection{Ejecta profile}\label{sec:hydro:prof}
\begin{figure}
 	 \includegraphics[width=\linewidth]{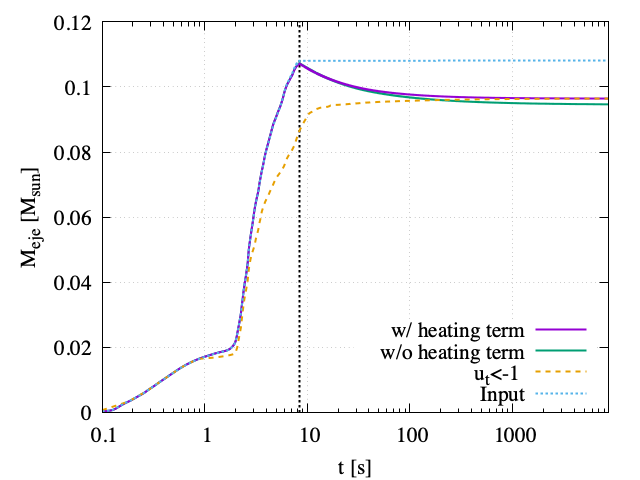}
 	 \caption{Total mass in the computational domain as a function of time. The purple and green curves denote the results for the HD simulations in which the radioactive heating is turned on and off, respectively. The orange dashed curve denotes the total mass of the material which is in the unbound trajectory. The blue dotted curve denotes the total input mass determined by the outflow data of the NR simulation ($u_t<-1$). The black dotted vertical line denotes the time at which the material injection from the inner boundary is truncated.}
	 \label{fig:meje_time}
\end{figure}

Figure~\ref{fig:meje_time} shows the time evolution of the total mass in the computational domain as well as the total input mass determined by the mass flux of the outflow data. As long as the material is injected from the inner boundary, the total mass in the computational domain agrees with the total input mass, and it finally reaches $\approx 0.11\,M_\odot$. As is discussed in~\cite{Fujibayashi:2020dvr}, two distinct mass ejection phases are seen: One found in the early phase ($t_{\rm in}\lesssim\,1$ s; referred to as the early-time ejecta component) which consists of the dynamical, early viscous driven and neutrino driven ejecta, and the other found in the late phase ($t_{\rm in}\gtrsim\,1$ s; referred to as the late-time ejecta component) which consists of the late-time viscous driven ejecta (see~\citealt{Fujibayashi:2020dvr} for the detail of each component). The first component contributes to $0.02\,M_\odot$ of the total ejecta mass, and the ejecta mass increases up to $\approx 0.11\,M_\odot$ by the contribution of the second component. After the outflow data from the NR result run out, a fraction of the material falls back through $r=r_{\rm in}$ and the total mass in the computational domain turns to decrease. The decrease rate becomes gradually smaller as the time evolves, and the total mass in the computational domain finally converges to $\approx0.096\,M_\odot$ for $t\geq100$ s. 

The fall back of the material happens because the pressure support from the inner boundary vanishes at the time when the outflow data run out. In~\cite{Fujibayashi:2020dvr}, whether the fluid element can escape from the system or not was judged by the Bernoulli's criterion, $hu_t<-h_{\rm min}\approx-0.9987\,c^2$, where $h$, $h_{\rm min}$, and $u_t$ are the specific enthalpy, the minimum value of $h$ in the EOS table employed (see Section 2.3 in~\citealt{Fujibayashi:2020dvr}), and lower time component of the four velocity, respectively. The material which reached the extraction radius in the NR simulation always satisfies this condition. On the other hand, by this definition, the fluid element counted as an ejecta component is not necessarily in a gravitationally unbound trajectory (i.e., which does not necessarily satisfy $u_t < -1$) at that time because the contribution from the internal energy is also taken into account for the ejecta determination. As long as the material is injected with sufficiently high pressure from the inner region, the fluid elements in the gravitationally bound orbits are gradually accelerated by the pressure gradient, and eventually transit to the unbound trajectories as in the stationary wind. However, after the pressure support from the inner boundary vanishes, the transition to the unbound trajectories is suppressed, and the material in the bound orbits is decelerated by the pressure from the preceding material. As a consequence, a fraction of the material falls back through $r=r_{\rm in}$. This suggests that the Bernoulli's criterion is not necessarily a sufficient condition for the material to be gravitationally unbound for a non-stationary flow (see also~\citealt{Kastaun:2014fna,Vincent:2019kor}).

To clarify this picture more quantitatively, we determine the material in the kinematically unbound trajectory by the fluid element which satisfies $u_t<-1$. Figure~\ref{fig:meje_time} shows that the total mass of the material in the unbound trajectories increases in time but with a slower rate than that of the total mass in the computational domain. The mass in the unbound trajectories reaches only up to $\approx0.093\,M_\odot$ at the time when the outflow data run out. The increase rate is significantly suppressed after the injection from the inner boundary is truncated, and eventually, the total mass in the computational domain converges to the value for the material in the unbound trajectories. The difference between the total mass in the computational domain and the total mass of the material in the unbound trajectories shows the mass of the material in the bound orbits. Thus, this indicates that the acceleration to a fraction of the injected material by the pressure gradient becomes inefficient after the time of the outflow truncation, and the material in the bound orbits falls back through the inner boundary. 

We should note that the fraction of the injected material which falls back depends on the epoch at which the pressure support from the inner region vanishes, and the sudden truncation of the injection from the inner boundary can be rather artificial. Nevertheless, we find that the final mass of the ejecta in the homologously expanding phase only varies by $\approx 10\%$ and some fraction of the injected material always falls back through the inner boundary even if the outflow injection is smoothly extended for a plausible time scale (the resulting light curves also do not significantly change; see Appendix~\ref{app:ext}). Thus, our results suggest that a fraction of the material which was counted as ejecta in the NR simulation in the late phase may fail to escape from the system.

\subsubsection{Effects of radioactive heating}\label{sec:hydro:heat}
\begin{figure*}
 	 \includegraphics[width=.5\linewidth]{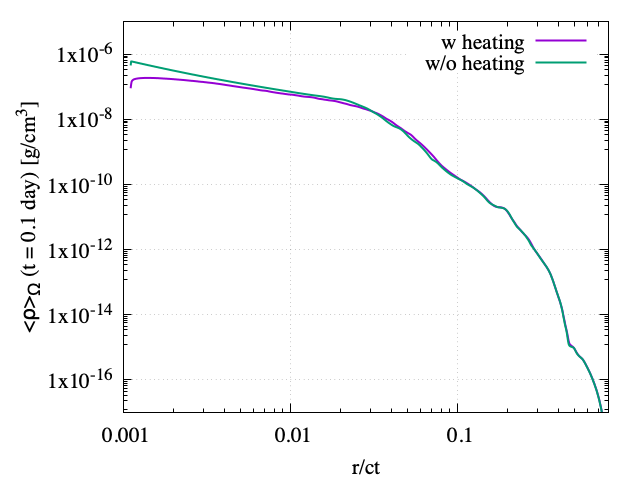}
 	 \includegraphics[width=.5\linewidth]{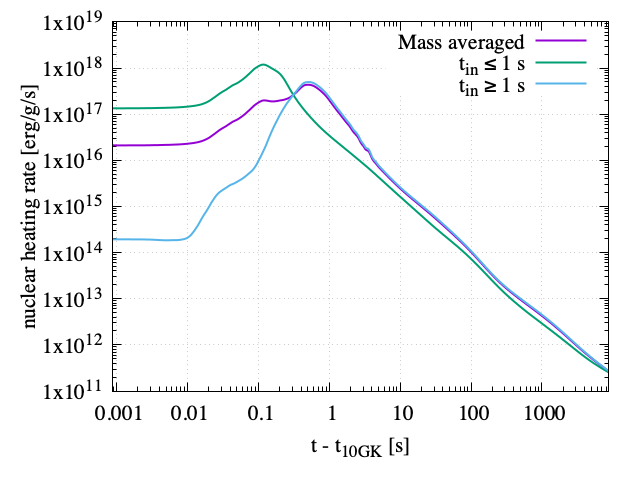}
 	 \caption{(Left panel) Angle-average rest-mass density of ejecta obtained for the snapshot at $t\approx 0.1\,{\rm days}$ as a function of radius with respect to $ct$. (Right panel) Specific radioactive heating rate of the ejecta. The origin of the horizontal axis is taken at the time when the fluid temperature decreased below $10\,{\rm GK}$. The purple, green, and blue curves denote the heating rates averaged among the whole ejecta, the ejecta injected before 1 s, and the ejecta injected after 1 s, respectively.}
	 \label{fig:meje_heat}
\end{figure*}

The result of our HD simulation in which the radioactive heating is turned off during the hydrodynamics evolution is also plotted in Figure~\ref{fig:meje_time} with the green curve. The time evolution of the total mass in the computational domain without the radioactive heating term agrees approximately with that for the fiducial setup, while the presence of the radioactive heating term slightly increases the total mass by enhancing the acceleration by the pressure gradient. This suggests that the radioactive heating plays only a minor role for the hydrodynamics evolution for our setup. Indeed, the left panel of Figure~\ref{fig:meje_heat} shows that the angle-averaged rest-mass density of the ejecta for the results with and without the radioactive heating term agrees approximately with each other. The presence of the radioactive heating term slightly decreases and increases the material with the radial velocity of $\leq 0.03\,c$ and $0.03$--$0.1\,c$, respectively, due to the enhancement of the pressure gradient, while the difference is nevertheless small. We note that the density at $r/ct\approx 10^{-3}$ for the model without the radioactive heating term is five times larger than that with the radioactive heating term, but the total mass in $r/ct \leq 0.005$ is less than $10^{-3}\,M_\odot$, and it is not important for kilonova lightcurves. On the other hand, the presence of the radioactive heating could induce a small difference in the resulting light curves due to the difference in the initial internal energy of the radiative transfer simulation (see Appendix~\ref{app:heat}).

The previous studies~\citep{Rosswog:2013kqa,Grossman:2013lqa} show that the radioactive heating could mildly modify the ejecta profile. In contrast to the previous results, the radioactive heating has a minor effect in our HD simulation. This can be understood by the delay until the ejected material reaches the extraction radius of the NR simulation. The right panel of Figure~\ref{fig:meje_heat} shows the specific radioactive heating rate of the ejecta as a function of time elapsed after the fluid temperature decreased below $10\,{\rm GK}$ (${\rm GK}=10^9{\rm K}$). While the contribution from radioactive heating rate becomes the largest at $t\approx1\,{\rm s}$, we find that the ejected material took typically $\approx 2\,{\rm s}$ until it reached the extraction radius of the NR simulation. Hence, the radioactive heating has been already weakened at the injection time in our HD simulation, and hence, has only a minor effect to the evolution.

However, we should note that the energy generation due to the nuclear reaction and $\beta$-decay in the NR simulation can be slightly underestimated. The EOS employed in the NR simulation considered the contribution from the binding energy of the nuclei in nuclear statistical equilibrium (NSE), and hence, the energy release due to the nuclear reaction is effectively taken into account as the release of the binding energy as long as NSE condition holds. However, as the ejecta expand, the temperature drops, and hence, the NSE is not established any longer. For such low temperature, the element abundances considered in the employed EOS are not very accurate. That is, for such a regime, the contribution from the energy generation due to, e.g., the {\it r}-process nucleosynthesis and successive radioactive decay, was not taken into account for the evolution of the ejecta in the NR simulation. Indeed, we find that the radioactive heating after the temperature drops below $\approx6$--$7$ GK can typically increases the internal energy by $\approx40\%$ at the time which the material reaches the extraction radius of the NR simulation. Nevertheless, by performing the long-term HD simulation even with the internal energy increased by $40\%$, we confirm that the acceleration of the ejecta velocity due to increase in the internal energy is within $\approx 5\%$, and the effect to the resulting light curves is only minor (see Appendix~\ref{app:intu14}).

\begin{figure*}
 	 \includegraphics[width=.5\linewidth]{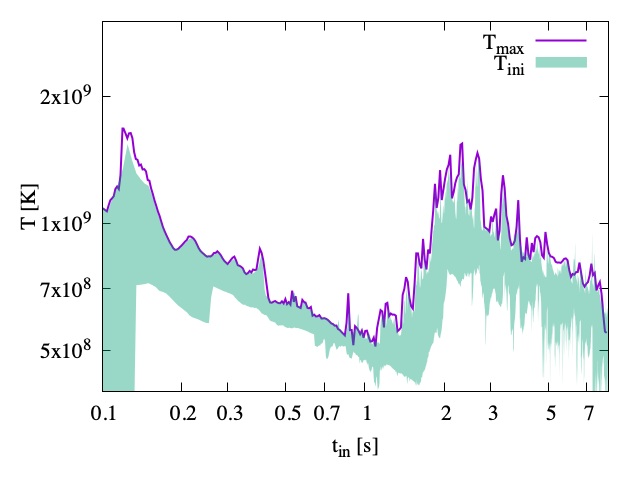}
 	 \includegraphics[width=.5\linewidth]{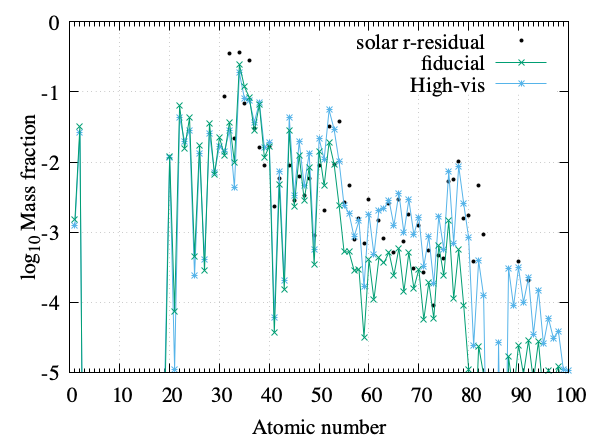}
 	 \caption{(Left panel) Maximum temperature which the material injected at $t=t_{\rm in}$ experienced during the HD simulation. The green shaded region denotes the initial temperature range of the ejected material at the time of injection. Note that the material which has fallen back into the inner boundary is not considered. (Right panel) Mass-averaged element mass fractions for the fiducial model (green lines) and the model with a large viscous parameter (blue lines, see Section~\ref{sec:LC:HVRes}) at $t=1$ day. The lanthanide mass fraction for the fiducial and higher viscosity models are $X_{\rm lan}\approx0.005$ and $X_{\rm lan}\approx0.03$, respectively. The solar {\it r}-residual pattern taken from~\cite{2020MNRAS.491.1832P} is plotted by the black points. Note that the solar {\it r}-residual pattern is shifted so that the Zr ($Z=40$) mass fraction agrees with that for the fiducial model.}
	 \label{fig:abuntemp}
\end{figure*}

The left panel of Figure~\ref{fig:abuntemp} shows the maximum temperature which the material injected at $t=t_{\rm in}$ experienced during the HD simulation. Here, assuming the domination of the radiation energy, temperature of the fluid is determined by $T=(e_{\rm int}/a)^{1/4}$ with $e_{\rm int}$ and $a$ being the internal energy density and the radiation constant, respectively. The range of the temperature at the time of injection is also shown by the green shaded region. The maximum temperature is always below $2\,{\rm GK}$. Note that the relatively high temperature is realized for the material with $t_{\rm in}\approx 0.1$ s and $t_{\rm in}\ge1$ s because their ejection is driven by shock heating and viscous heating, respectively. The small deviation of the maximum temperature from the high edge of the initial temperature range indicates that the fluid element does not experience significant heat up during the hydrodynamics evolution for $r>8000$ km. Note that, in~\cite{Fujibayashi:2020dvr}, the thermodynamic histories of tracer particles were extrapolated for nucleosynthesis calculations up to about 1 year by assuming the constant terminal velocity and entropy. Our result indicates that the hydrodynamics evolution after the material escaped from the computational domain of the NR simulation has only a minor effect on the nucleosynthesis.

The right panel of Figure~\ref{fig:abuntemp} shows the mass-averaged element mass fractions for the fiducial model and the model with a large viscous parameter (see Section~\ref{sec:LC:HVRes}) at $t=1$ day. The lanthanide mass fractions of the fiducial and higher viscosity models are $X_{\rm lan}\approx0.005$ and $X_{\rm lan}\approx0.03$, respectively. While the detail discussion can be found in~\cite{Fujibayashi:2020dvr}, we emphasize here that, for both models, a large amount of the first {\it r}-process peak elements is present in the ejecta. As we see in Section~\ref{sec:LC:Res:fid}, the first {\it r}-process peak elements have a large impact on the resulting kilonova light curves.

\subsubsection{2D profile}\label{sec:hydro:2dprof}
\begin{figure}
 	 \includegraphics[width=.9\linewidth]{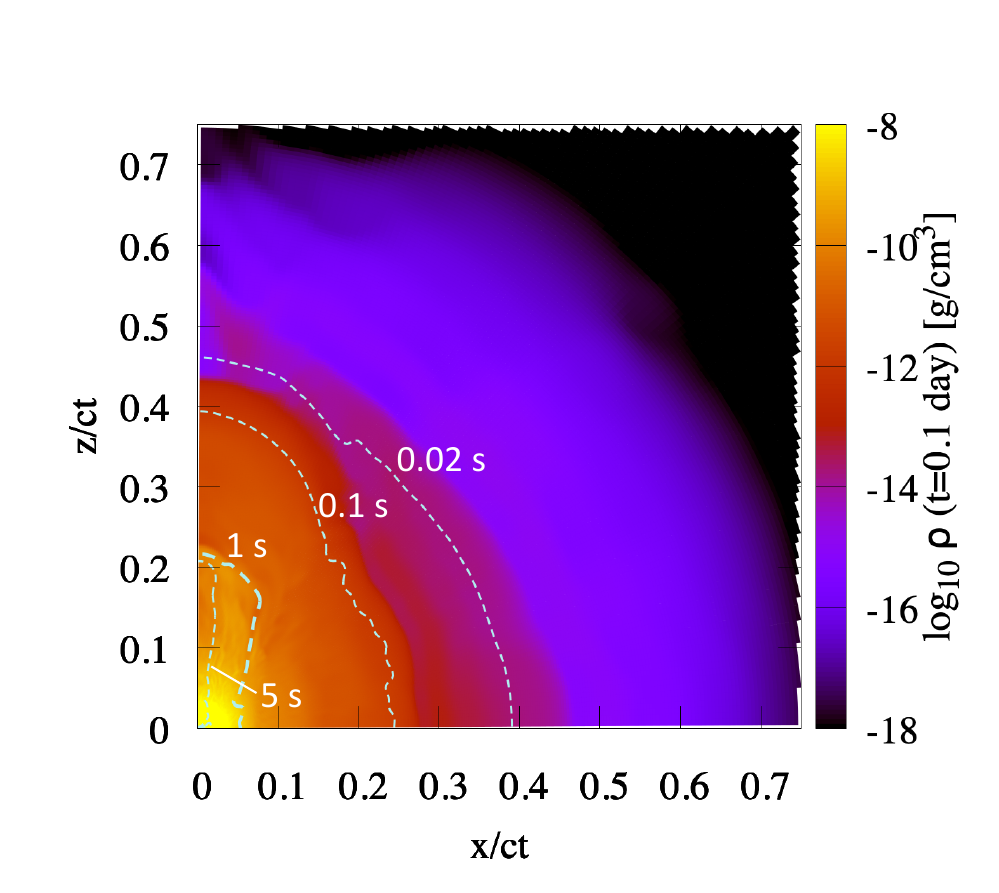}
 	 \includegraphics[width=.9\linewidth]{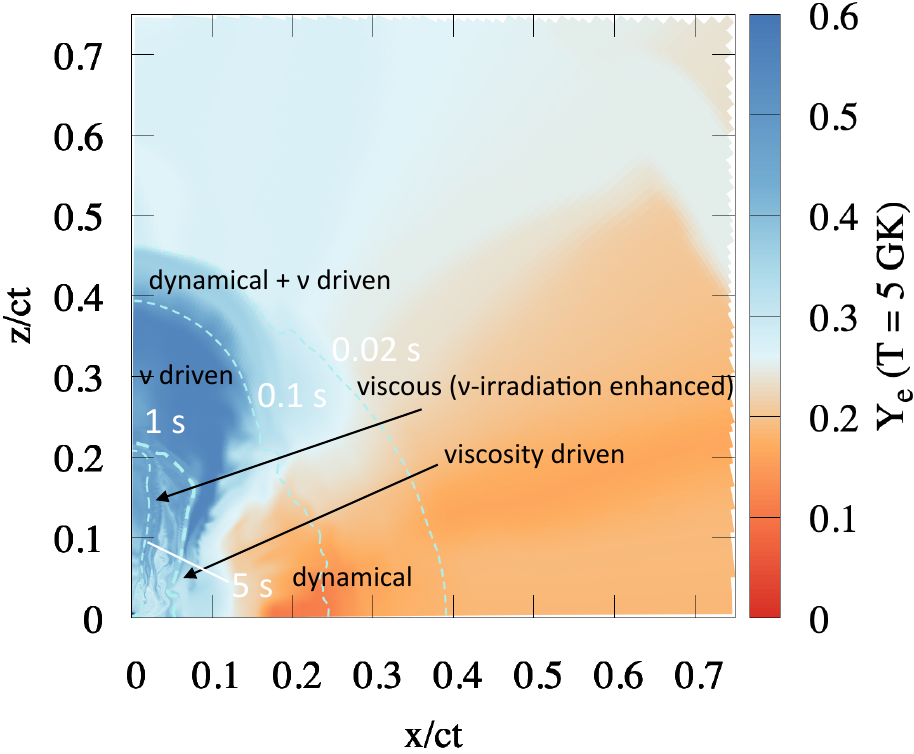}
 	 \includegraphics[width=.9\linewidth]{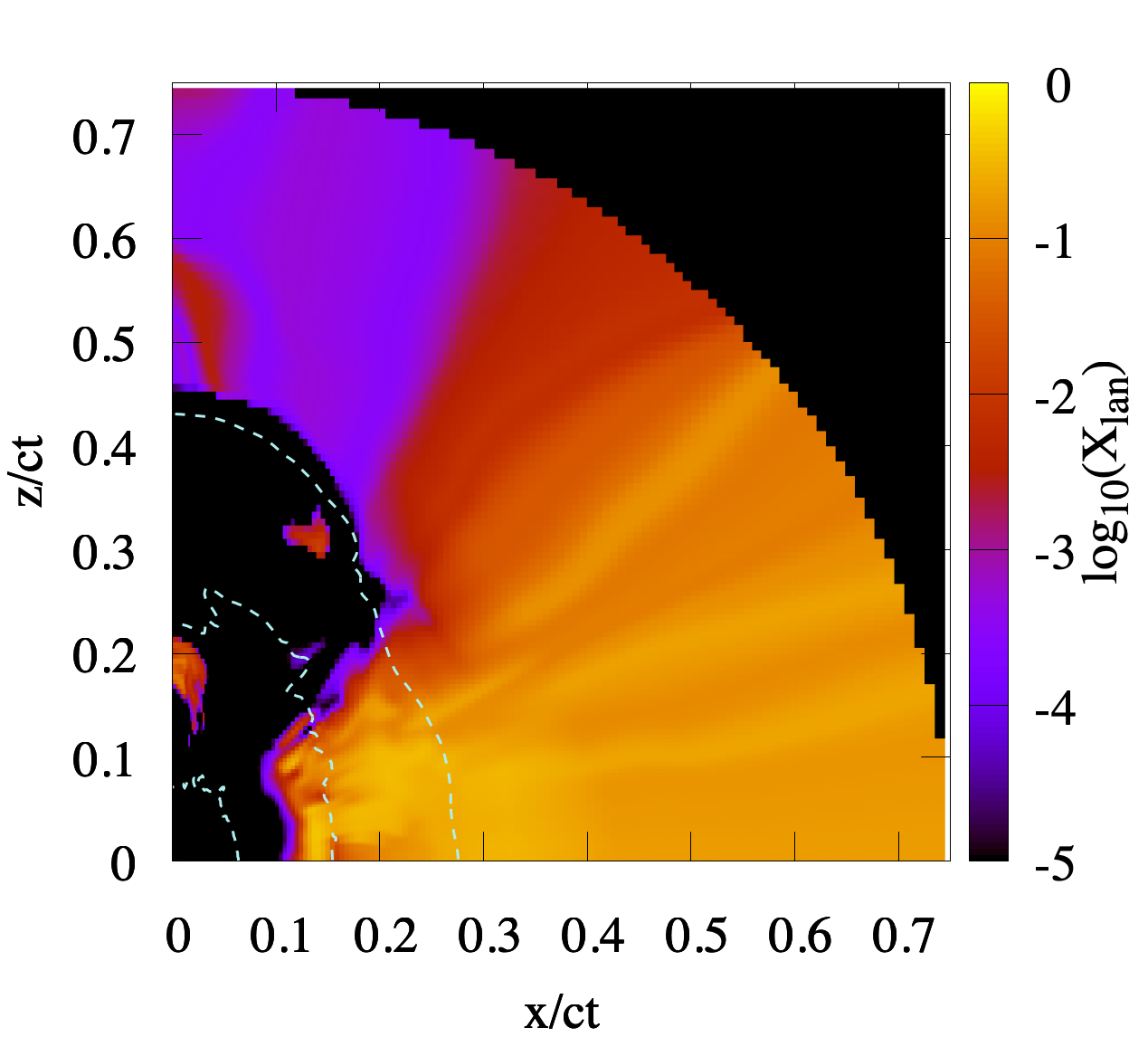}
 	 \caption{(Top panel) Rest-mass density profile of the ejecta obtained by the HD simulation at $t\approx 0.1\,{\rm days}$. The dashed curves denote the ejecta which escape from the extraction radius of the NR simulation at $t=0.02$, $0.1$, $1$, and $5\,{\rm s}$. (Middle panel) $Y_{\rm e}$ value at the temperature of 5 GK for each fluid element in the ejecta profile at $t\approx 0.1\,{\rm days}$. The typical launching mechanism for each part of ejecta is described by the black-colored label. (Bottom panel) Distribution of the lanthanide ($57\le Z\le71$) mass fraction at 1 day with $Z$ being the atomic number. The dashed curves denote the rest-mass density contour of $(10^{-13},\,10^{-11},\,10^{-9})\,{\rm g/cm^3}$ at $t=0.1$ day from the outside.}
	 \label{fig:prof_dens}
\end{figure}

The top and middle panels of Figure~\ref{fig:prof_dens}, respectively, show the rest-mass density and $Y_{\rm e}$ value at the temperature of 5 GK for each fluid element in the ejecta profile at $t\approx 0.1\,{\rm days}$. As is shown in Appendix~\ref{app:homo}, the ejecta are already approximately in the homologously expanding phase at $t\approx 0.1\,{\rm day}$, and thus, $r/ct$ approximately shows the velocity of the fluid. 

Both early ($t_{\rm in}\lesssim1$ s) and late ($t_{\rm in}\gtrsim1$ s) time ejecta components exhibit mildly prolate morphology. The former component ($t_{\rm in}\lesssim1$ s) distributes from $\approx0.1\,c$ to $0.3\,c$ for the equatorial region and from $\approx0.2\,c$ to $0.4\,c$ for the polar region. The latter component ($t_{\rm in}\gtrsim1$ s) exhibits more elongated shape, and it is entirely surrounded by the former component. The higher ejecta velocity in the polar direction, which is the origin of the prolate shape, is due to neutrino irradiation from the remnant NS. Indeed, Figure 5 in~\cite{Fujibayashi:2020dvr} shows that the ejecta velocity is enhanced in the presence of neutrino irradiation. This indicates that the ejecta from a BNS merger that results in a long-lived remnant NS can always have a prolate shape.

Discontinuity in the density distribution is found for the material injected between $ \approx 0.02$ s and $ \approx 0.1$ s (the edge of the dark-orange region in the top panel of Figure~\ref{fig:prof_dens} located from $(x/ct,z/ct)\approx(0,0.43)$ to $(x/ct,z/ct)\approx(0.28,0)$) and for the material in the polar region injected at $ \approx 1$ s ($(x/ct,z/ct)\approx(0,0.23)$). The discontinuity found for the material of $t_{\rm in}\approx 0.1$ s indicates the presence of the interaction between the early viscous or the neutrino driven ejecta component and the low velocity edge of the preceding dynamical ejecta component, while the discontinuity found for the material of $t_{\rm in}\approx 1$ s is formed between the late-time viscous and the early viscous/neutrino driven ejecta components.

The $Y_{\rm e}$ profile of the early-time ejecta component ($t_{\rm in}\lesssim1$ s) shows a clear angular dependence. For $\theta\lesssim\pi/4$, the value of $Y_{\rm e}$ is always above 0.3. In particular, the early-time ejecta component is dominated by the material with $Y_{\rm e}>0.4$ for $\theta\lesssim\pi/6$. On the other hand, the material with $Y_{\rm e}<0.3$ dominates the early-time ejecta component with the radial velocity larger than $\approx0.1\,c$ for $\theta\gtrsim\pi/4$. Compared to the early-time ejecta component, the $Y_{\rm e}$ value of the late-time ejecta component ($t_{\rm in}\gtrsim1$ s) shows less spatial dependence, and it is always approximately in a range of $0.3$--$0.5$.

The bottom panel of Figure~\ref{fig:prof_dens} shows the distribution of the lanthanide ($57\le Z\le71$) mass fraction of the ejecta profile at 1 day, where $Z$ denotes the atomic number. Reflecting the angular dependence of the $Y_{\rm e}$ profile, a large value of the lanthanide fraction is realized in the equatorial region of the early-time ejecta component. On the other hand, only a tiny amount of lanthanides is synthesized in the polar region of the early-time ejecta component ($\theta\lesssim\pi/4$) and in almost the entire region of the late-time ejecta component as a consequence of the high values of $Y_{\rm e}$. As we show in the next section, the presence of lanthanides in the equatorial region of the early-time ejecta component has a great impact on the resulting kilonova light curves.

\begin{figure}
 	 \includegraphics[width=\linewidth]{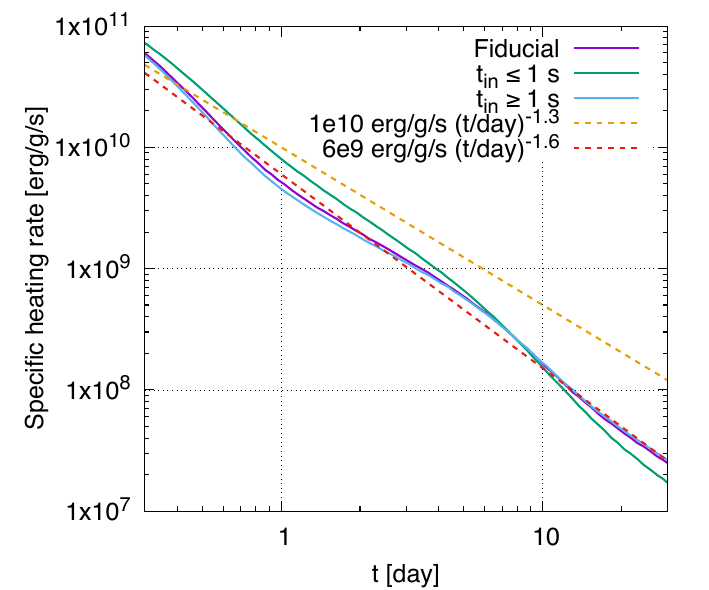}
 	 \caption{Specific heating rate of the entire ejecta and the early/late-time ejecta components for $t=0.3$--$30$ days. The thermalization efficiency is taken into account by employing an analytic formula of~\cite{Barnes:2016umi}.}
	 \label{fig:nuc_fid}
\end{figure}

Figure~\ref{fig:nuc_fid} shows the specific heating rate of the entire ejecta and the early/late-time ejecta components for $t=0.3$--$30$ days, which is responsible for the kilonova emission. Here, the thermalization efficiency is taken into account by employing an analytic formula (Equations (32)—(33)) of~\cite{Barnes:2016umi} for $\gamma$-rays, electrons, $\alpha$-particles, and fission fragments. The specific heating rate of the entire ejecta agrees approximately with that of the late-time ejecta component ($t_{\rm in}\geq1$ s). It can be approximated by $6\times10^{9}\,{\rm erg/g/s}\,(t/ 1\,{\rm day})^{-1.6}$ for $t\approx0.3$--30 days, which shows much steeper time dependence than the heating rate typically realized for ejecta with $Y_{\rm e}<0.25$, that is, $\approx10^{10}\,{\rm erg/g/s}\,(t/ 1\,{\rm day})^{-1.3}$~\citep[e.g.,][]{Metzger:2010sy,Wanajo:2014wha}. At $t\approx1$ day, the late-time ejecta component has a specific heating rate lower approximately by a factor of 2 than that of the early-time ejecta component ($t_{\rm in}\leq1$ s). This reflects that $r$-process nucleosynthesis takes place only weakly in the late-time ejecta due to the high values of $Y_{\rm e}$~\citep{Wanajo:2014wha,Wu:2016pnw,Lippuner:2017bfm}. For $t\geq10$ days, the specific heating rate of the early-time ejecta component becomes lower than that of the late-time ejecta component because the thermalization efficiency is lower for the early-time ejecta component due to the lower rest-mass density.

\section{Light curves}\label{sec:LC}
\subsection{Radiative transfer simulation}\label{sec:LC:RT}
We calculate the light curves from the obtained ejecta profile using a wavelength-dependent radiative transfer simulation code~\citep{Tanaka:2013ana,Tanaka:2017qxj,Tanaka:2017lxb,Kawaguchi:2019nju}. The photon transfer is calculated by a Monte Carlo method for given ejecta profiles of the density, velocity, and element abundance. We also consider the time-dependent thermalization efficiency following an analytic formula derived by~\cite{Barnes:2016umi}. Axisymmetry is imposed for the matter profile, such as the density, temperature, and abundance distribution. The ionization and excitation states are calculated under the assumption of local thermodynamic equilibrium (LTE) by using the Saha ionization and Boltzmann excitation equations. 

For photon-matter interaction, we consider bound-bound, bound-free, and free-free transitions and electron scattering for a transfer of optical and infrared photons~\citep{Tanaka:2013ana,Tanaka:2017qxj,Tanaka:2017lxb}. The formalism of the expansion opacity~\citep{1983ApJ...272..259F,1993ApJ...412..731E,Kasen:2006ce} and the updated line list calculated in~\cite{Tanaka:2019iqp} are employed for the bound-bound transitions. The line list is constructed by an atomic structure calculation for the elements from $Z=26$ to $Z=92$, and supplemented by Kurucz's line list for $Z < 26$~\citep{1995all..book.....K}. Note that, since our atomic data include only up to the triple ionization for all the ions, the early parts of the light curves ($t\le 1\,{\rm days}$) may not be very reliable due to too high ejecta temperature, while the error in the {\it ugrizJHK}-band light curves for $0.5\,{\rm days} \le t\le 1\,{\rm days}$ is expected to be $\lesssim0.5$ mag (see~\citealt{Banerjee:2020myd} for the work taking the opacity contribution from higher ionization states into account).

The radiative transfer simulations are performed from $t=0.1\,{\rm days}$ to $30\,{\rm days}$. We employ the snapshot of our HD simulation at $0.1\,{\rm days}$ and the density profile is mapped to the velocity space in the Cartesian coordinates assuming the homologous expansion. The initial internal energy and temperature for the radiative transfer simulations are also determined from those obtained by our HD simulation. The spatial distributions of the heating rate and element abundances are determined by the table obtained by the nucleosynthesis calculations referring to the injected time and angle of the fluid elements. In particular, we employ element abundances at $1\,{\rm days}$ and fix them during the time evolution to reduce the computational cost. We checked that this prescription has only a minor effect on the results by performing the same radiative transfer simulation but employing element abundances at $10\,{\rm days}$.
\subsection{Results}\label{sec:LC:Res}
\subsubsection{Fiducial model}\label{sec:LC:Res:fid}
\begin{figure}
 	 \includegraphics[width=1.\linewidth]{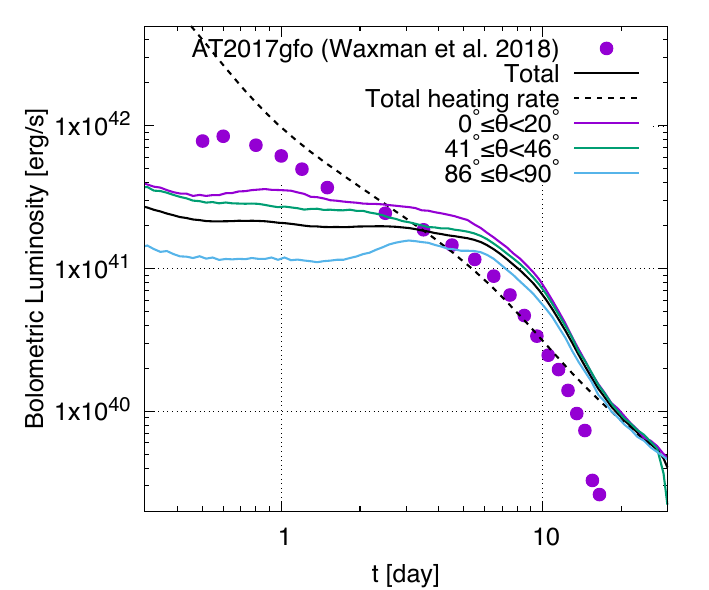}
 	 \caption{Bolometric luminosity of the fiducial model. The isotropic bolometric luminosity for various viewing angles is also shown. The dashed curve and the data points denote the total (thermalized) heating rate and the estimated bolometric luminosity of AT2017gfo taken from~\cite{Waxman:2017sqv} assuming the distance of $40$ Mpc, respectively.}
	 \label{fig:lbol_fid}
\end{figure}

Figure~\ref{fig:lbol_fid} shows the bolometric luminosity of the fiducial model. The luminosity remains nearly constant until $t\approx3$ days and steeply declines at $t\approx7$ days. The bolometric luminosity observed from the polar and equatorial direction is larger and smaller by a factor of $\approx 2$ than the viewing angle averaged value at $t=1\,{\rm day}$, respectively. This is due to the blocking effect by the lanthanide-rich ejecta located around the equatorial plane and resulting preferential diffusion of photons in the polar direction~\citep{Kasen:2014toa,Kawaguchi:2019nju,Bulla:2019muo,Zhu:2020inc,Darbha:2020lhz,Korobkin:2020spe}. The viewing angle dependence vanishes after $\approx 10$ days, which suggests that the entire ejecta are optically thin for such a phase.

As pointed out in~\cite{Fujibayashi:2020dvr}, the total heating rate of the fiducial model exhibits a similar shape to the bolometric luminosity measured in AT2017gfo assuming the distance of 40 Mpc. On the other hand, we find that the bolometric luminosity actually realized by taking the photon diffusion effect into account shows different features from that observed in AT2017gfo; the bolometric luminosity is lower by more than a factor of 2 at $t=1$ day, while the luminosity for $t\ge3\,{\rm days}$ is higher than the observation, irrespective of the viewing angles. The suppression of the bolometric luminosity from the total heating rate for $t\lesssim3$ days is due to the trapping of photons by the optically thick media, while the overshooting of the luminosity from the instantaneous heating rate for $3\,{\rm days}\lesssim t\lesssim15\,{\rm days}$ is due to the release of the trapped radiation energy as a consequence of the density decrease. The bolometric luminosity agrees with the instantaneous heating rate after the entire ejecta become optically thin and the trapped radiation energy is all released ($t\gtrsim15\,{\rm days}$).

The fainter emission in the early phase ($\approx 1$ days) indicates that the diffusion time scale of photons is longer than in AT2017gfo. The brighter emission in the late phase ($\approx 7$ days) indicates that the total ejecta mass of AT2017gfo are smaller than that of the fiducial ejecta model ($\approx0.1\,M_\odot$) unless the uncertainty of the thermalization efficiency is significantly large. Thus, this result supports that the progenitor of AT2017gfo is not likely to be a system like our fiducial model, that is, a BNS merger that results in a long-lived remnant NS.

\begin{figure*}
 	 \includegraphics[width=.5\linewidth]{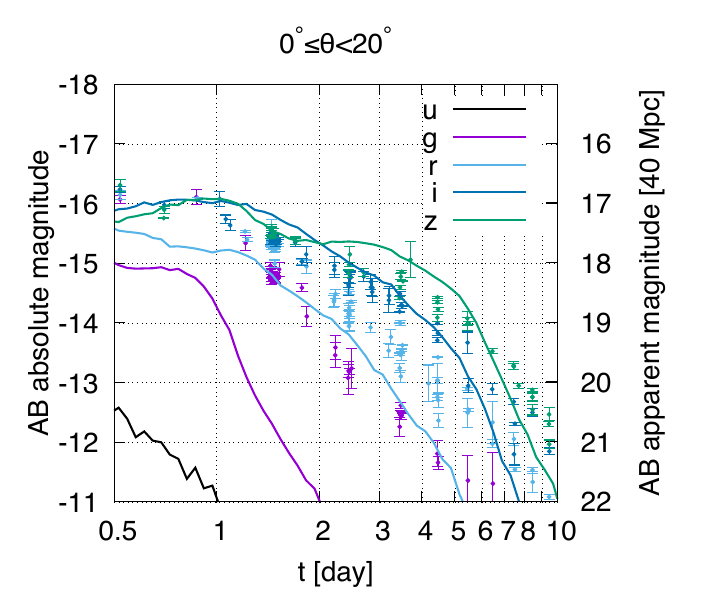}
 	 \includegraphics[width=.5\linewidth]{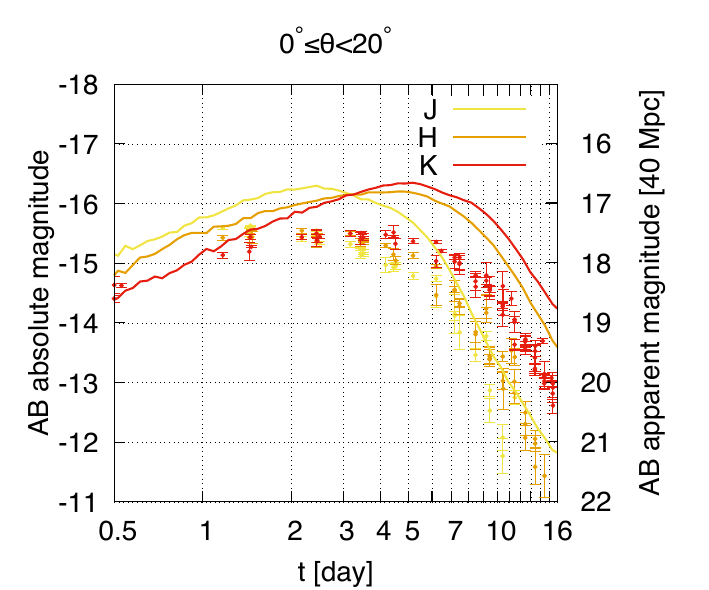}\\
 	 \includegraphics[width=.5\linewidth]{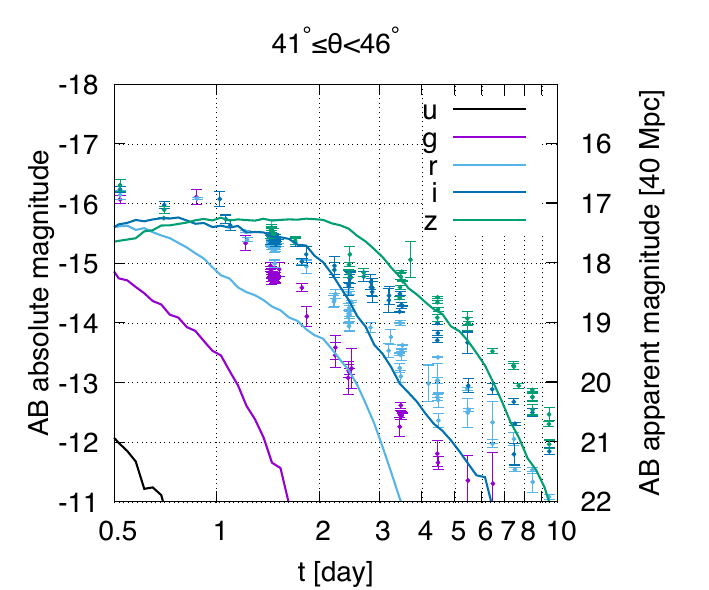}
 	 \includegraphics[width=.5\linewidth]{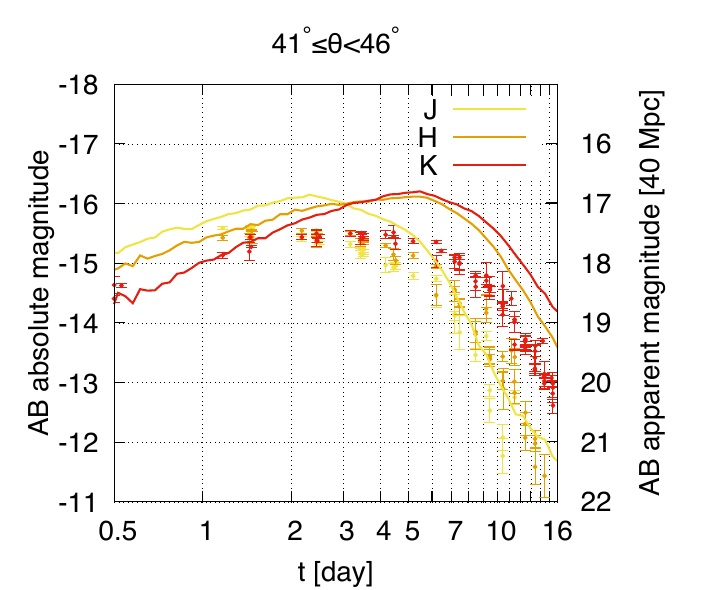}\\
 	 \includegraphics[width=.5\linewidth]{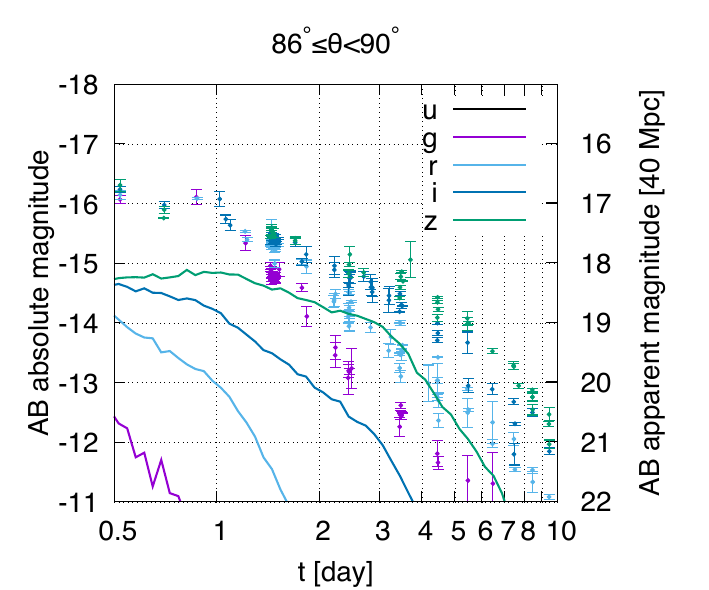}
 	 \includegraphics[width=.5\linewidth]{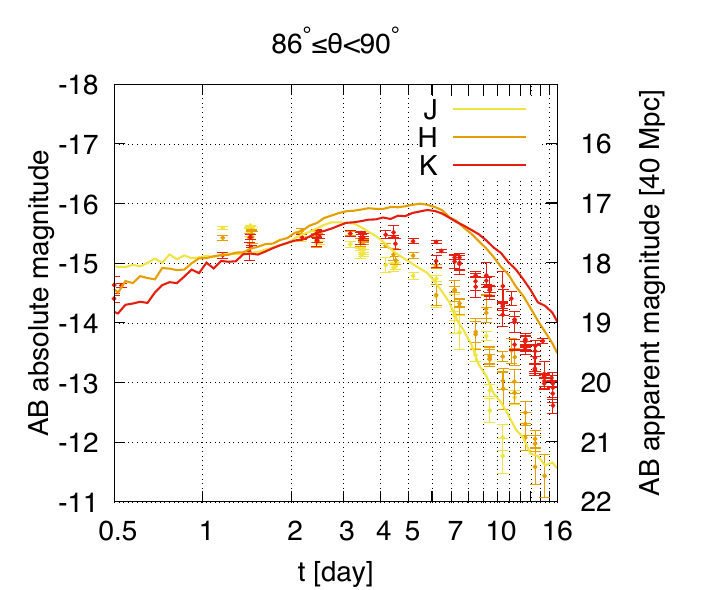}
 	 \caption{The {\it ugriz} (the left panels) and {\it JHK}-band (the right panels) light curves observed from $0^\circ\le\theta\le20^\circ$ (the top panels), $40^\circ\le\theta\le46^\circ$(the middle panels), and $86^\circ\le\theta\le90^\circ$ (the bottom panels). Data points denote the observed data of AT2017gfo summarized in~\cite{Villar:2017wcc}. The hypothetical distance of $40$ Mpc is used for the apparent magnitudes.}
	 \label{fig:mag_fid}
\end{figure*}

Figure~\ref{fig:mag_fid} shows the {\it ugrizJHK}-band light curves observed from various viewing angles. Despite the low lanthanide fraction of the ejecta model, we find that the kilonova light curves for the fiducial model are relatively faint in the optical wavelengths and rather bright in the infrared wavelengths compared to AT2017gfo or the low lanthanide fraction models in the previous studies~\cite[e.g., {\tt HMNS\_YH} in][]{Kawaguchi:2019nju}. Indeed, while the emission in the {\it riz}-band observed from $0^\circ\leq\theta\leq20^\circ$ for $t\geq 1$ day agrees broadly with those observed in AT2017gfo assuming the distance of $40\,{\rm Mpc}$, we find that the emission in the {\it g}-band is fainter than those observed in AT2017gfo. Instead, the emission in the {\it JHK}-band for $t\geq 1$ day is brighter by $\approx0.5$--$1\,{\rm mag}$ than those observed in AT2017gfo.

\begin{figure*}
 	 \includegraphics[width=.5\linewidth]{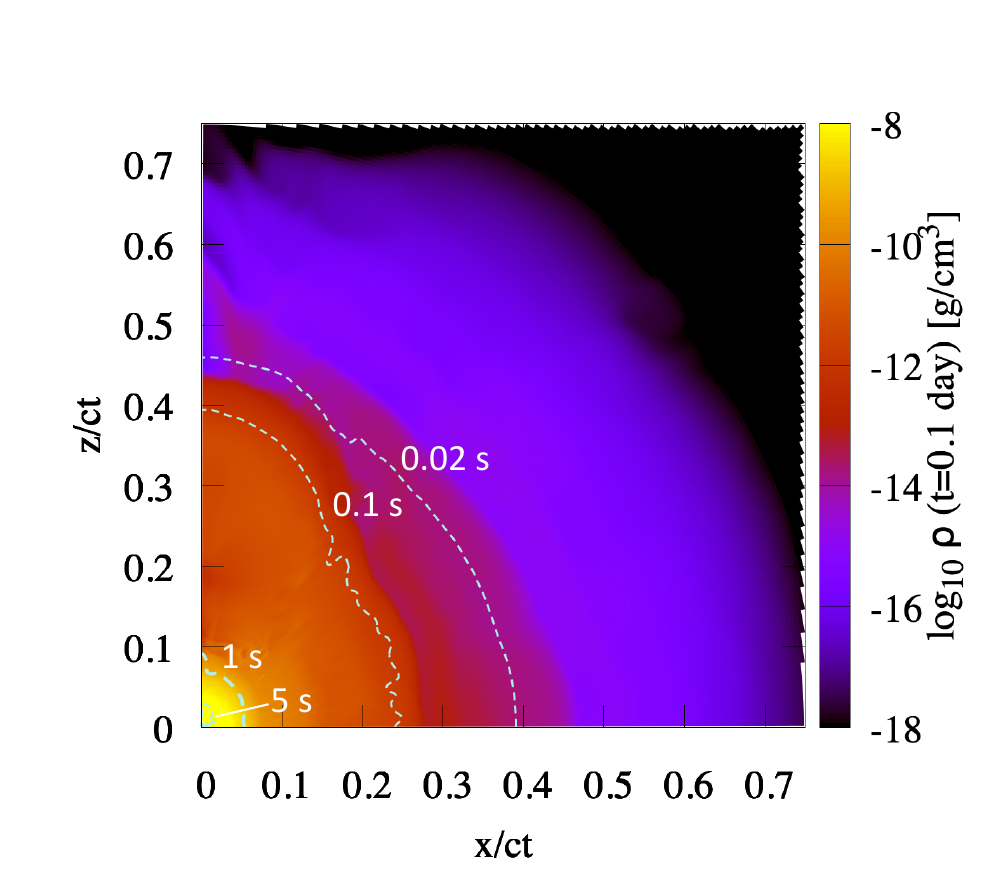}
	 \includegraphics[width=.5\linewidth]{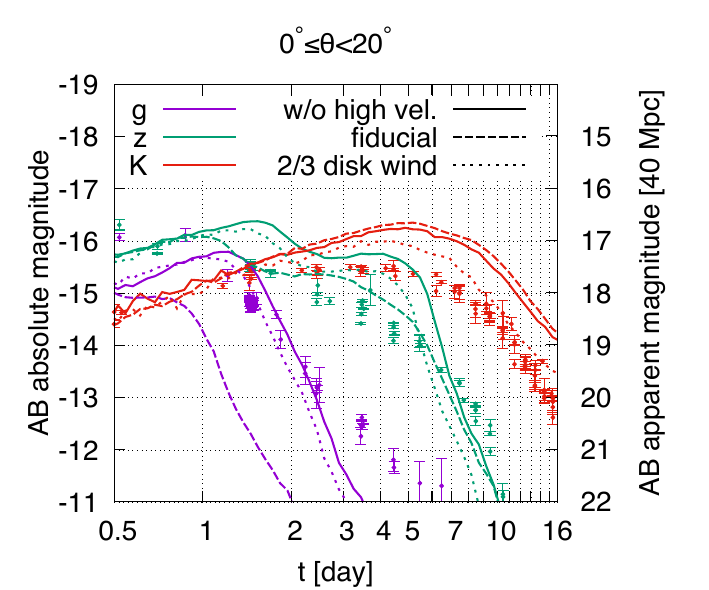}
 	 \caption{(Left panel) Rest-mass density profile of the ejecta at $t\approx 0.1\,{\rm days}$ for the HD simulation in which the outflow injection from $\theta\le 30^\circ$ is truncated for $t\ge 200$ ms in order to suppress the high-velocity late-time ejecta component in the polar region ($v_z\approx0.1$--0.2 $c$). The dashed curves denote the ejecta which escape from the extraction radius of the NR simulation at $t=0.02$, $0.1$, $1$, and $5\,{\rm s}$. (Right panel) Comparison of the {\it gzK}-band light curves observed from $0^\circ\le\theta\le20^\circ$ between the fiducial ejecta model (the dashed curves) and the ejecta model in which the high-velocity late-time ejecta component is suppressed (``w/o high vel.", the solid curves). The dotted curves (``2/3 disk wind") denote the same as ``w/o high vel." but for the case the outflow mass for $\theta\ge 30^\circ$ is also reduced to 2/3 of the original value for $t\ge 200$ ms. The meaning of the data points is the same as in Figure~\ref{fig:mag_fid}.}
	 \label{fig:magcomp_jet0}
\end{figure*}

There are three reasons that cause the faint optical and bright infrared emission. The first reason is in the prolate morphology of the late-time ejecta component (see Figure~\ref{fig:prof_dens}). For the prolate morphology, photons diffuse preferentially toward the equatorial direction, in which the optical depth is small~\citep{Tanaka:2013ixa,Kyutoku:2015gda,Kawaguchi:2019nju,Zhu:2020inc,Darbha:2020lhz,Korobkin:2020spe}. In such a situation, optical photons emitted from the late-time ejecta component are efficiently reprocessed into infrared photons in the lanthanide-rich early-time ejecta component located around the equatorial plane. 

To investigate the effect of the prolate morphology, we performed a HD simulation in which the outflow injection from $\theta\le 30^\circ$ is truncated for $t\ge 200$ ms. By this prescription, the high-velocity late-time ejecta component in the polar region (with $z/ct=0.1$--$0.2$; see the contour of $t_{\rm in}=1$ s) is suppressed. The left panel of Figure~\ref{fig:magcomp_jet0} shows that the rest-mass density profile of the late-time ejecta component ($t_{\rm in}\ge 1$ s) exhibits approximately a spherical morphology. For this case, as shown in the right panel of Figure~\ref{fig:magcomp_jet0}, we find that the optical emission is indeed enhanced. Note that broadly the same results are obtained for different truncation time of the polar outflow ($100$--$400$ ms). In the right panel of Figure~\ref{fig:magcomp_jet0}, we also show the results for the case that the outflow mass for $\theta\ge 30^\circ$ is also reduced to 2/3 of the original value, which might mimic a BNS merger that accompanies the formation of a black hole (see the discussion on this in Section~\ref{sec:D}).

We note that not only the morphology but also the density structure is also the key for the effect explained above. For example,~\cite{Korobkin:2020spe} studied the effect of the ejecta morphology in the presence of the multiple ejecta components. Among the models studied in~\cite{Korobkin:2020spe}, the models with torus-shaped low-$Y_{\rm e}$ ejecta and and peanut-shaped high-$Y_{\rm e}$ ejecta are most similar to our fiducial model. However, their models with such morphology show the bolometric luminosity comparable to the observation of AT2017gfo for $\leq1\,{\rm day}$ whereas our model results in smaller luminosity than the observation despite the larger ejecta mass (see, in particular, "T1P1" or "T2P2" in Table 2 of in~\citealt{Korobkin:2020spe}). We speculate that, while the difference in the element abundances and heating rate might be the main cause for this difference, the difference in the density structure can also be responsible for it: While the peanut-shaped high-$Y_{\rm e}$ ejecta model of~\cite{Korobkin:2020spe} has the density peak in $(x/ct,z/ct)\approx(0,0.1$--$0.2)$, the late-time ejecta component ($t_{\rm in}\geq 1\,{\rm s}$) of our model is more concentrated in the center region $(r/ct<0.1)$. Such difference in the density profile causes large difference in the diffusion time scale of photons, and hence, in the resulting bolometric luminosity (e.g.,~\citealt{Kawaguchi:2019nju}). Moreover, the presence of the relatively smaller density region in the polar region ($z/ct=0.1$--$0.2$) in our model also has a great impact to suppress the ultraviolet and optical lightcurves (e.g., see the extended Figure 1 in~\citealt{Kasen:2017sxr}). Thus, we should consider not only the ejecta morphology but also the density structure as the key characteristics of the ejecta.


\begin{figure*}
 	 \includegraphics[width=.5\linewidth]{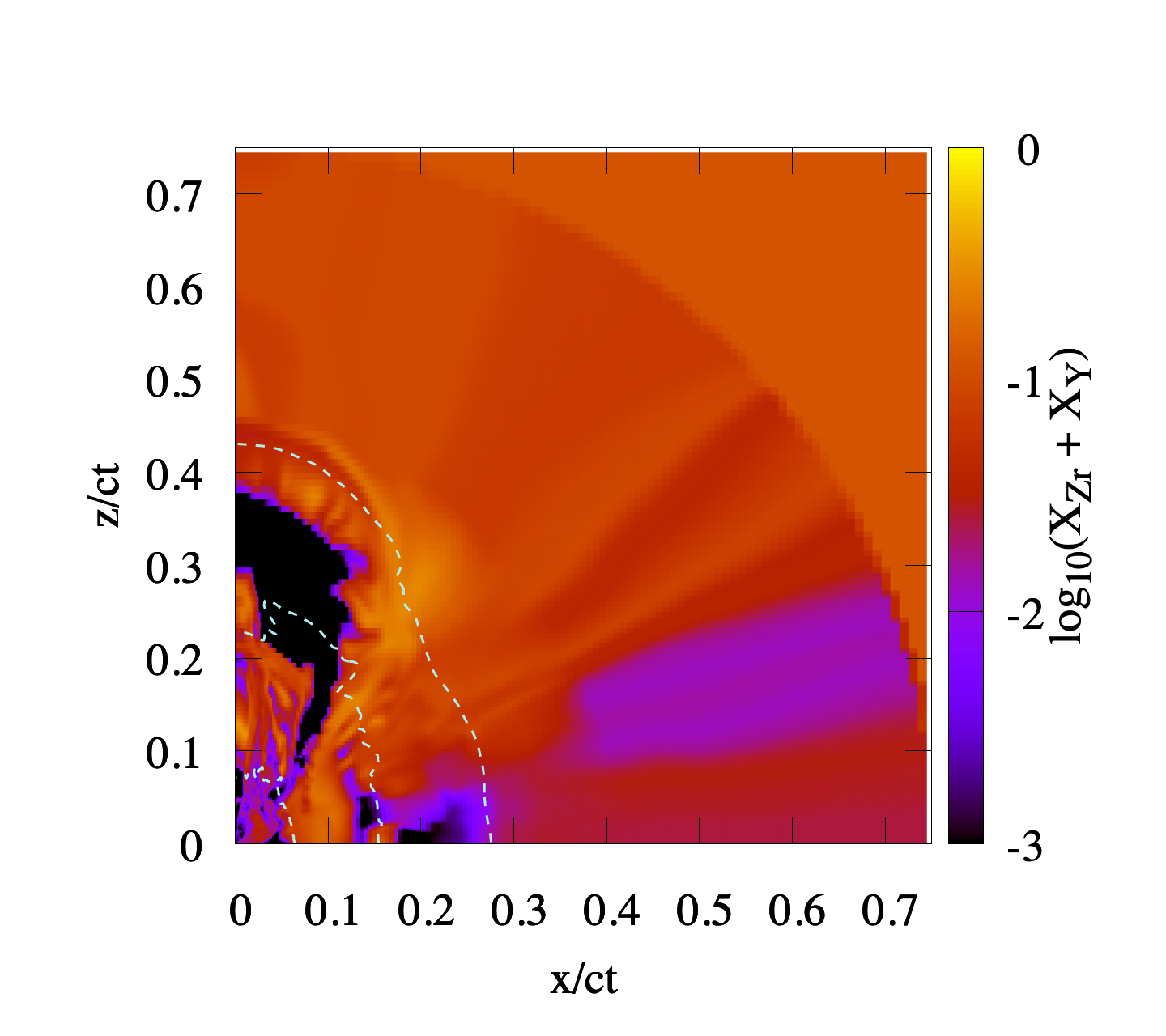}
 	 \includegraphics[width=.5\linewidth]{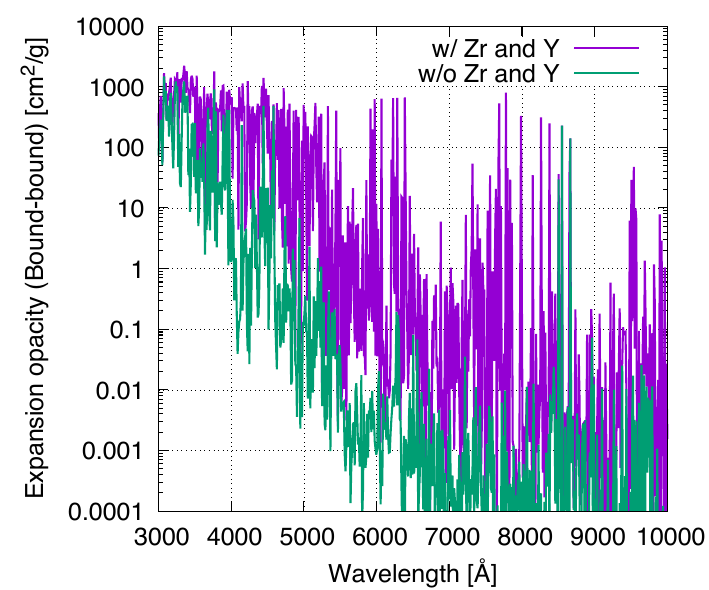}
 	 \caption{(Left panel) Distribution of the Y ($Z=39$) and Zr ($Z=40$) mass fraction at 1 day. The dashed curves denote the rest-mass density contour of $(10^{-13},\,10^{-11},\,10^{-9})\,{\rm g/cm^3}$ at $t=0.1$ day from the outside. (Right panel) Comparison of the expansion opacity~\citep{1983ApJ...272..259F,1993ApJ...412..731E,Kasen:2006ce} of bound-bound transitions between the cases that the contributions from Y and Zr are taken (the blue curve) and not taken into account (the green curve). The opacity is calculated by employing the element abundances in $(x/ct,z/ct)=(0.001,\,0.4)$ and by assuming $2\times 10^{-16}\,{\rm g/cm^3}$ and $3000\,$K, which correspond to the condition at $t=\,1.5$ days.}
	 \label{fig:ZrY}
\end{figure*}

\begin{figure*}
 	 \includegraphics[width=.5\linewidth]{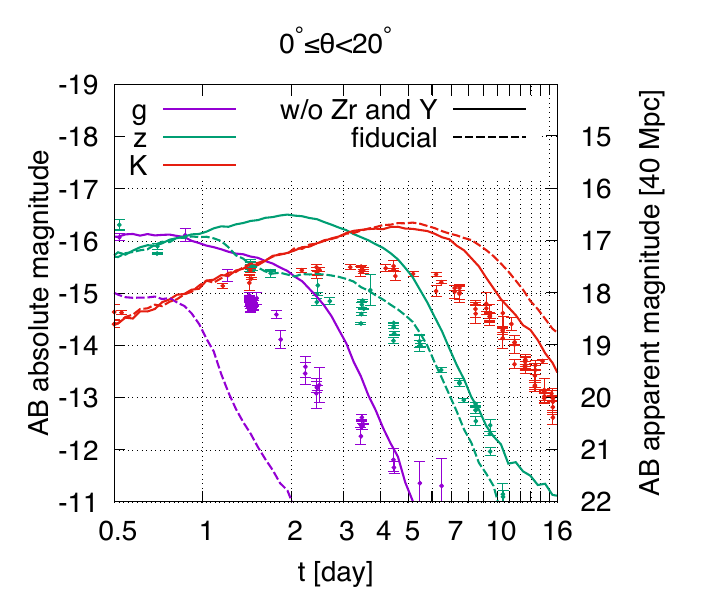}
	 \includegraphics[width=.5\linewidth]{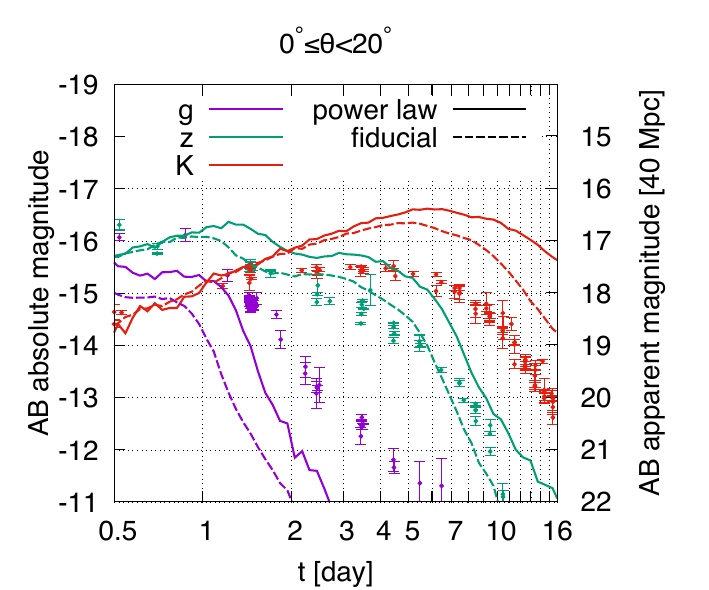}
 	 \caption{(Left panel) Comparison of the {\it gzK}-band light curves observed from $0^\circ\le\theta\le20^\circ$ between the fiducial setup (the dashed curves) and the case in which the opacity contributions from Y and Zr are omitted (the solid curves). (Right panel) Comparison of the {\it gzK}-band light curves observed from $0^\circ\le\theta\le20^\circ$ between the fiducial setup (the dashed curves) and a hypothetical case with the power law heating rate of $10^{10}\,{\rm erg/g/s}\,(t/ 1\,{\rm day})^{-1.3}$ (the solid curves). The meaning of the data points is the same as in Figure~\ref{fig:mag_fid}.}
	 \label{fig:magcomp_2nd3rd}
\end{figure*}

The second reason is in the presence of Y ($Z=39$) and Zr ($Z=40$) in the high velocity edge of the ejecta. Y and Zr are categorized as {\it d}-shell elements, and they significantly contribute to the opacity in the optical wavelength as they have a large number of low-lying energy levels~\citep{Tanaka:2019iqp}. These elements (and also Sr) are abundantly produced in the ejecta with relatively high values of $Y_{\rm e}$ (e.g., $>$ 0.3; ~\citealt{Wanajo:2014wha}). The left panel of Figure~\ref{fig:ZrY} shows the distribution of the Y ($Z=39$) and Zr ($Z=40$) mass fraction at 1 day. Focusing on the polar region ($\theta\lesssim 30^\circ$), a large amount of Y and Zr as well as 
relatively low-mass $r$-process elements is present in the high velocity edge of the early and late-time ejecta components with $z/ct\approx$ 0.2 and $0.4$, respectively. In the right panel of Figure~\ref{fig:ZrY}, we compare the expansion opacity~\citep{1983ApJ...272..259F,1993ApJ...412..731E,Kasen:2006ce} of bound-bound transitions between the cases that the contributions from Y and Zr are taken and not taken into account. Here, the opacity is calculated by employing the element abundances in $(x/ct,z/ct)=(0.001,\,0.4)$ and by assuming the density of $2\times 10^{-16}\,{\rm g/cm^3}$ and temperature of $3000\,$K, which correspond to the condition there at $t=\,1.5$ days. It shows that the bound-bound opacity in the polar edge of the ejecta is dominated by the contribution from Y and Zr. 

The left panel of Figure~\ref{fig:magcomp_2nd3rd} compares the {\it gzK}-band light curves observed from $0^\circ\le\theta\le20^\circ$ between the fiducial setup and the case in which the opacity from Y and Zr is omitted. As is shown in the figure, the opacity from Y and Zr is responsible for suppressing the {\it ugr}-band emission by $\approx1$--2$\,{\rm mag}$ for $t\gtrsim1\,{\rm day}$. We note that Y and Zr are also present in the equatorial region, but the opacity is dominated by the lanthanide elements there. We also note that the first $r$-process peak elements including Y and Zr are less produced in the polar region of the ejecta with $0.3\lesssim r/ct \lesssim0.4$ due to a large $Y_{\rm e}$ value ($\ge 0.5$).

The third reason is in the relatively low specific heating rate of the late-time ejecta component (see Figure~\ref{fig:nuc_fid}). The mass averaged heating rate of the ejecta in the fiducial setup is typically lower than that for the ejecta with $Y_{\rm e}<0.25$ by a factor of $2$--$4$ for $t\approx1$--$10$ days. The low heating rate leads to low ejecta temperature, and hence, shifts the spectral peak to the longer wavelengths. The right panel of Figure~\ref{fig:magcomp_2nd3rd} shows that the emission is by $\approx0.5$--$1$ mag dimmer than that in a hypothetical model with $10^{10}\,{\rm erg/g/s}\,(t/ 1\,{\rm day})^{-1.3}$ for $t\approx1$--$10$ days that reasonably approximates that of ejecta with $Y_{\rm e}<0.25$~\citep[e.g.,][]{Metzger:2010sy,Wanajo:2014wha}.

\subsubsection{Higher viscosity model}\label{sec:LC:HVRes}
\begin{figure*}
 	 \includegraphics[width=.5\linewidth]{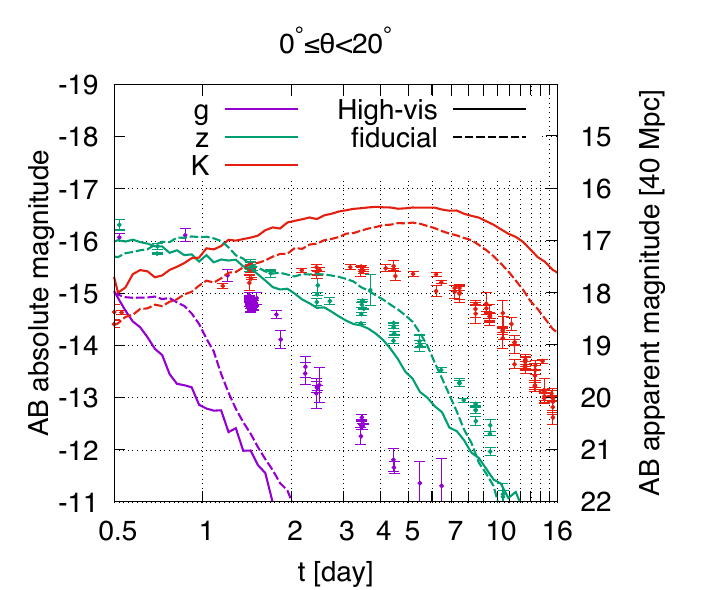}
	 \includegraphics[width=.5\linewidth]{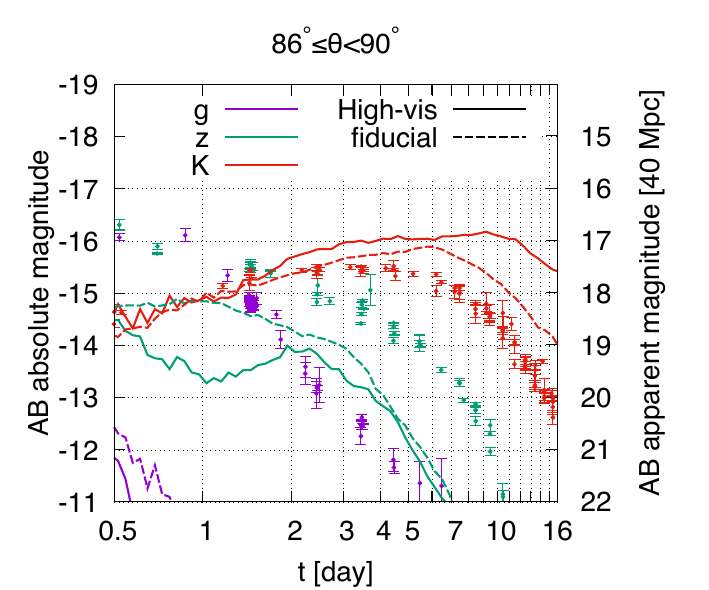}
 	 \caption{Comparison of the {\it gzK}-band light curves between the fiducial ejecta model (the dashed curves) and the ejecta model employing the outflow data of the NR simulation with a large viscous parameter (the solid curves). The light curves observed from $0^\circ\le\theta\le20^\circ$ (the left panel) and $86^\circ\le\theta\le90^\circ$ (the right panel) are shown. The meaning of the data points is the same as in Figure~\ref{fig:mag_fid}.}
	 \label{fig:magcomp_v10}
\end{figure*}

To investigate how the mass ejection time scale of the accretion disk surrounding the remnant NS changes the resulting kilonova light curves, we perform a long-term HD simulation and a radiative transfer simulation for the outflow data obtained in the NR simulation with a large viscous parameter (DD2-125M-h) in the same manner as for the fiducial model (see Appendix~\ref{app:HV} for the resulting ejecta profile). Figure~\ref{fig:magcomp_v10} compares the {\it gzK}-band light curves for the fiducial ejecta model and for the ejecta model with a large viscous parameter. 

We find that the light curves for the model with a large viscous parameter show the features similar to those for the fiducial model, that is, the relatively faint optical and bright infrared emission. This is due to the rest-mass density and element abundance distributions of the higher viscosity model similar to those of the fiducial model: the prolate shape of the late-time ejecta component with high mass of $\approx 0.13\,M_\odot$, torus-like distribution of lanthanide elements in the surrounding early-time ejecta component, and presence of the first {\it r}-process peak elements in the polar region. 

Our result may indicate that the kilonova emission from BNS mergers that result in long-lived remnant NSs share broadly common features despite the difference in the nucleosynthesis yields if the viscous parameter is in the plausible range  ($\lesssim 0.1$; e.g., Kiuchi et al. 2018; Fernández et al. 2019). Note that, in a quantitative manner, the optical and infrared emission are fainter and brighter by $\approx1$ mag, respectively, than those for the fiducial model because the late-time ejecta component of the higher viscosity model has a larger lanthanide fraction~\citep[see the right panel of Figure~\ref{fig:abuntemp} and e.g.,][]{Kawaguchi:2019nju}.

We note, however, that the NR simulation with a higher viscous parameter was performed for the limited simulation time ($\lesssim2$ s), and a large fraction of the ejected material ($\approx0.05\,M_\odot$) still remained inside the extraction radius. Such an ejecta component is neglected in our long-term HD simulation since the injection of the outflow is truncated after the end of the outflow data. We should note that the final ejecta mass in the homologously expanding phase will be larger, and hence, the emission brighter than that found in Figure~\ref{fig:magcomp_v10} can be realized particularly in the infrared wavelengths if such an ejecta component is taken into account in our long-term HD simulation.

\section{Discussion}\label{sec:D}
\begin{figure*}
 	 \includegraphics[width=.5\linewidth]{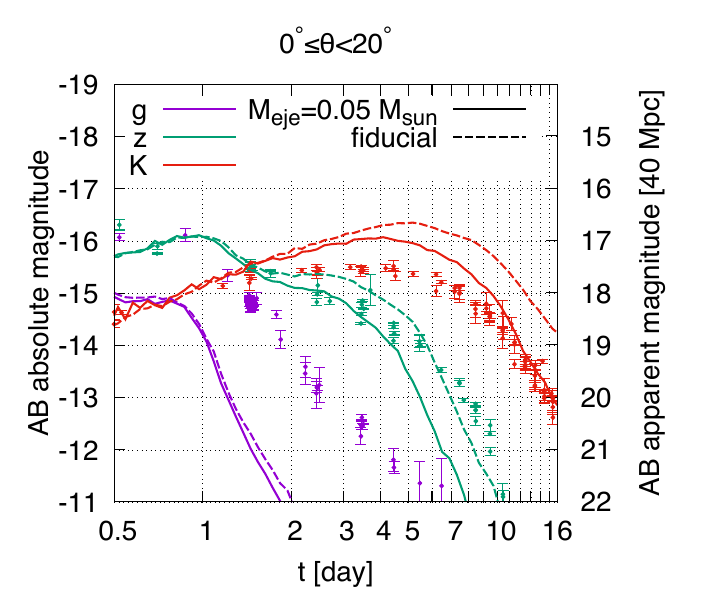}
	 \includegraphics[width=.5\linewidth]{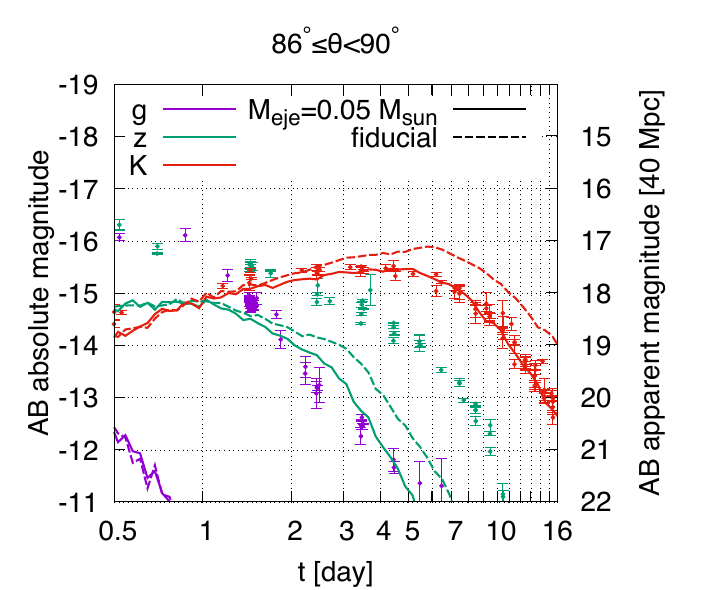}
 	 \caption{Comparison of the {\it gzK}-band light curves between the fiducial ejecta model (the dashed curves) and the model in which the outflow injection is truncated after $t_{\rm in}=3$ s and the total ejecta mass is reduced by a factor of $\approx2$ ($\approx0.05\,M_\odot$; the solid curves). The light curves observed from $0^\circ\le\theta\le20^\circ$ (the left panel) and $86^\circ\le\theta\le90^\circ$ (the right panel) are shown. The meaning of the data points is the same as in Figure~\ref{fig:mag_fid}.}
\label{fig:magcomp_var}
\end{figure*}
~\cite{Fujibayashi:2020dvr} showed that BNS mergers that accompany the formation of long-lived remnant NSs will result in qualitatively the similar ejecta outcome irrespective of the mass of each NS and the EOS for the NS. Quantitatively, there is a variation in the total ejecta mass up to a factor of 2 among the models with different EOS and different total binary mass reflecting difference in the remnant torus mass (see Table 3 in~\citealt{Fujibayashi:2020dvr}). Nevertheless, the property of the kilonova lightcurves for the models with different EOS and different total mass can be qualitatively the same as for the fiducial model. In Figure~\ref{fig:magcomp_var}, we compare the {\it ugrizJHK}-band lightcurves between the fiducial model and the model in which the outflow injection is truncated after $t_{\rm in}=3$ s and the total ejecta mass is reduced by a factor of $\approx2$. Note that the latter model approximately corresponds to the model with SFHo (SFHo-125H in ~\citealt{Fujibayashi:2020dvr}). Figure~\ref{fig:magcomp_var} shows that the kilonova lightcurves have qualitatively the same property as for the fiducial model for $t=1$--16 days, that is, relatively faint optical and bright infrared emission.

We also found that the kilonova light curves with qualitatively similar features can be realized even for the case with a large viscous parameter, with which material with more heavy nuclei is ejected. We speculate that the lightcurves with qualitatively the similar feature can also be obtained for the case with a smaller viscous parameter: If the viscous parameter is smaller than that employed in the fiducial model, the mass ejection time scale becomes longer. For such a case, the post-merger ejecta with a slightly higher value of $Y_{\rm e}$ will be formed, and a smaller amount of heavy nuclei will be synthesized in the ejecta~\citep{Fujibayashi:2020qda,Fujibayashi:2020jfr,Fujibayashi:2020dvr}. On the other hand, the presence of the first $r$-process peak elements in the polar edge ($z/ct\approx 0.4$) and the lanthanides in the equatorial region, which are the important opacity sources, is expected to be less dependent to the viscous parameter as they are originated by the dynamical and neutrino driven mass ejection (see the middle panel of Figure~\ref{fig:prof_dens}). Since the radioactive heating rate in the ejecta will be strongly suppressed due to suppressed production of the radioactive heavy nuclei, the brightness of the optical emission for the case with a smaller viscous parameter could be comparable to or even smaller than the fiducial model. Thus, while performing the NR simulations with a smaller viscous parameter is necessary for the confirmation, our results imply that the mergers of BNS systems that accompany the formation of long-lived remnant NSs may always result in similar kilonova light curves to what we found in this work.

~\cite{Fujibayashi:2020dvr} pointed out that equal-mass BNS mergers that result in long-lived remnant NSs will not be the main events among the entire mergers, as the resulting nucleosynthetic yields are different from the solar $r$-process-like pattern which is observed in the $r$-process-enhanced metal-poor stars~\citep[e.g.,][]{Cowan:2019pkx}. Nevertheless, BNS mergers that result in long-lived remnant NSs may be detected in the future, as a low-mass NS binary indeed exists~\citep{Stovall:2018ouw}. Our results show that, if the event similar to the fiducial model occurs, the kilonova emission will be intrinsically brighter than $-14$ mag in the {\it z}-band for $1\leq t\leq3$ days irrespective of the viewing angles. Such emission is detectable by the observation using 1-m class and 4/8-m class telescopes if the distance to the event is within $\sim$100 Mpc and 200 Mpc, respectively~\citep{Nissanke:2012dj}. The infrared observation employing the telescopes, such as VISTA~\citep{Ackley:2020qkz}, will further increase the detectability of the kilonovae since the {\it HK}-band emission is apparently brighter than 21 mag for $1\leq t\leq10$ days for all the viewing angles if the distance to the event is within $\sim$150 Mpc. On the other hand, to detect the optical emission of which wavelengths are shorter than the {\it g}-band, the follow-up observation within $\approx 1$ day is crucial unless the event occurs much closer than the case of AT2017gfo. The discovery of a kilonova from such a system will be a good opportunity to test our current understanding of the merger process and emission mechanism.

The relatively faint optical emission found in our kilonova model is partly due to the presence of a large amount of material in the polar region. The abundant existence of Y and Zr in this region also plays a role for darkening the optical polar emission. This indicates that the suppression of the high-velocity ejecta components in the polar region with a larger amount of the first {\it r}-process peak elements may be needed for the kilonova to be as bright in the optical wavelength as in AT2017gfo. Indeed, we found that the optical emission will be enhanced and the light curves could be broadly consistent with AT2017gfo for the case that the high velocity component in the polar region is suppressed (see Figure~\ref{fig:magcomp_jet0}).

Since the high velocity components in the polar region is enhanced by neutrino irradiation from the long-lived remnant NS, the high-velocity ejecta components in the polar region could be suppressed if the remnant NS collapses into a black hole in a shorter time scale after the merger. For such a case, the viscous driven disk outflow will also be suppressed by a factor of $\approx1/3$--$2/3$ \citep{Fujibayashi:2020qda,Fujibayashi:2020jfr}. Figure 8 shows that the infrared light curves approach those of AT2017gfo for the case that the mass of the viscous driven ejecta is reduced to 2/3 of the original value. (see ``2/3 disk wind" in Figure~\ref{fig:magcomp_jet0}). Note that the probable identification of Sr in the spectra of AT2017gfo~\citep{2019Natur.574..497W} implies that the ejecta as a whole of this BNS merger should also contain a substantial amount of Y and Zr (that are co-produced with Sr). This suggests that the presence of bright optical emission in a kilonova may indicate the collapse of the remnant NS to a black hole in a short time scale of ${\cal O}(0.1)$\,s (see also~\citealt{Mosta:2020hlh}). 

However, we should note that the results could be modified if the magneto-hydrodynamical effects are taken into account in the NR simulations. As discussed in~\cite{Fujibayashi:2020dvr}, the remnant NS and disk can be strongly magnetized by magneto-hydrodynamical process such as Kelvin-Helmholtz instability~\citep{Price:2006fi,Kiuchi:2017zzg}, the magneto-rotational instability~\citep{Balbus:1998ja}, and the winding effects. In such a case, ejecta can be accelerated more efficiently by the magneto-hydrodynamical effects than those in the prescription of effective viscosity. In fact, it is shown in~\cite{Mosta:2020hlh} that the ejecta velocity is significantly enhanced up to $0.5\,c$ by magneto-hydrodynamical process in the presence of a strong poloidal field (see also~\citealt{Fernandez:2018kax,Ciolfi:2020wfx,Shibata:2021bbj}). Thus, the ejecta can be more extended in the velocity space if the magneto-hydrodynamical effects are taken into account.

Our speculation to the kilonova lightcurves in the presence of magnetically driven high-velocity post-merger ejecta components is as follows. If the ejecta formation occurs in a short time scale ($\lesssim 0.1$ s),  $Y_{\rm e}$ of ejecta will be low and a significant amount of lanthanides might be synthesized~\citep{Fujibayashi:2020qda,Fujibayashi:2020jfr,Fujibayashi:2020dvr}. For such a case, the outer-most layer of the ejecta can be lanthanide-rich, and the kilonova lightcurves might be much fainter and brighter in the optical and infrared wavelengths, respectively, due to the so-called lanthanide curtain effect~\citep{Kasen:2014toa}. On the other hand, if the mass ejection time scale is $\gtrsim 1$ s,  $Y_{\rm e}$ will be high enough so that the post-merger ejecta are lanthanide-poor as in the fiducial model of this work. For such a case, kilonova lightcurves brighter in the optical wavelengths for the early phase ($\approx 1$ day) might be realized due to high ejecta velocity (e.g.,~\citealt{Kawaguchi:2019nju}). However, the other important thermodynamical condition of ejecta, such as specific entropy of the ejecta, can also be significantly altered in the presence of strong magnetically driven turbulence, and thus, it is not trivial what kinds of nuclear/element abundances, which determine the opacity and the heating rate, are realized in the ejecta. Thus, the study based on the NR simulations taking the effects of magneto-hydrodynamics is crucial. 

Moreover, as discussed in~\cite{Nativi:2020moj} and~\cite{Klion:2020efn}, the interaction between the preceding ejecta and the relativistic jet launched from the central remnant black hole-torus system may also enhance the optical emission by blowing off the ejected material with the first $r$-process peak elements including Y and Zr located in the polar region. For such a case, the collapse of the remnant NS to a black hole in a short time scale of ${\cal O}(0.1)$\,s might not be necessarily needed to interpret the observation of AT2017gfo. In fact, some studies~\citep[e.g.,][]{Murguia-Berthier:2020tfs} suggest that the post-merger mass ejection lasts for $\sim1$ s can be consistent with the afterglow emission observed in GW170817. Further systematic investigation based on the NR simulations with the collapsing remnant NSs taking the effects of magneto-hydrodynamics and relativistic jets into account is needed to get deeper understanding of the event and integrating the knowledge obtained by various aspects.

Finally, we discuss a possible non-LTE effect which can modify the emission for $t\gtrsim1$ day. In our radiative transfer simulation, we assume LTE and the ionization state population of the atoms are determined by solving Saha's equation~\citep{Lucy:2004fz,Tanaka:2013ana}. However, this assumption could break down for the low density region in which the recombination rates of ions become smaller than the ionization rates. Indeed, the importance for taking the non-LTE effect in the excitation/ionization population into account is well known for supernovae radiative transfer simulations~\citep[e.g.,][]{Boyle:2016zcr}. Moreover,~\cite{Hotokezaka:2021ofe} suggest that the population of neutral atoms can be significantly suppressed in the nebula phase due to the suppression of the recombination rate as the consequence of the density drop.

\begin{figure*}
 	 \includegraphics[width=.5\linewidth]{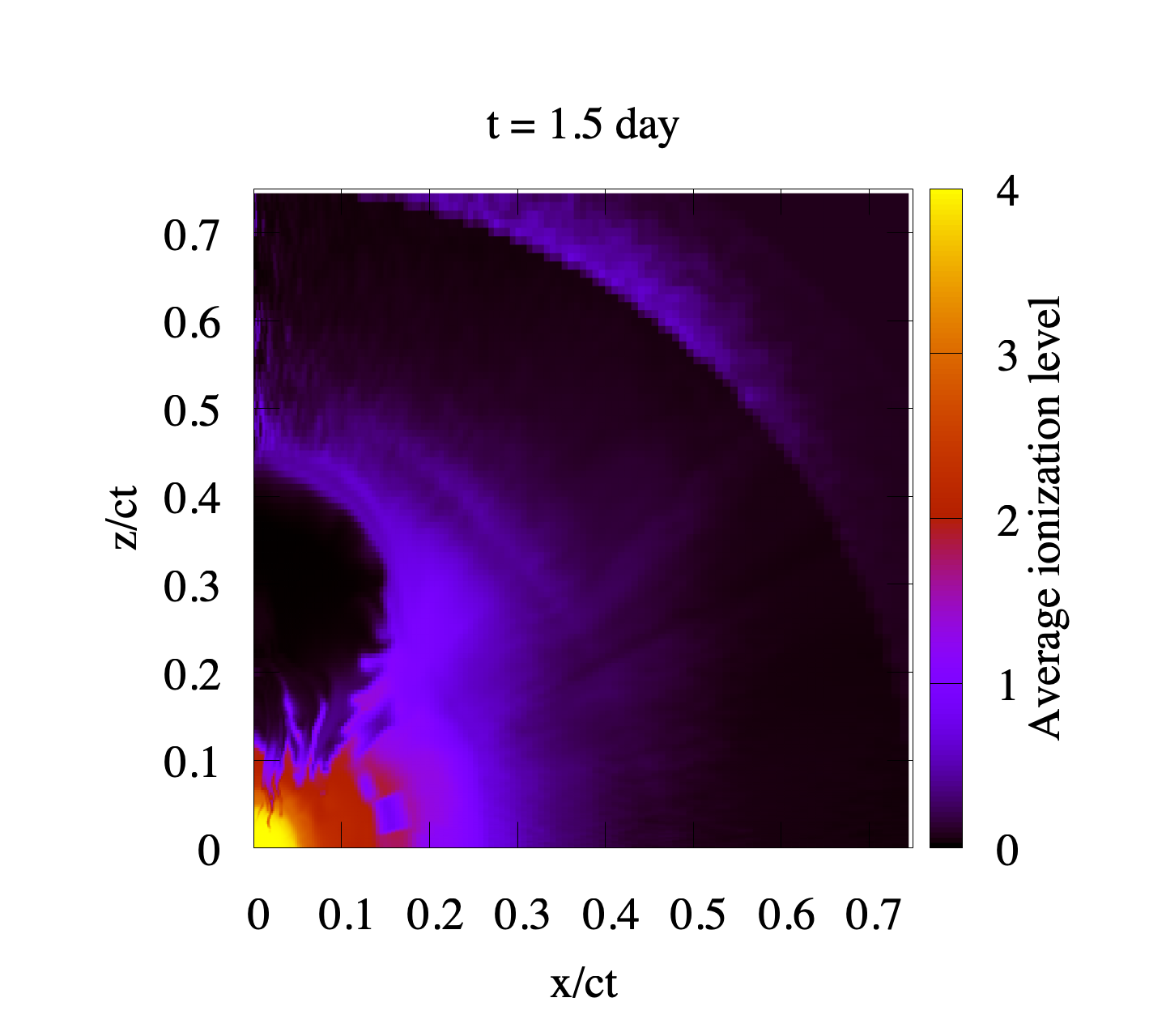}
 	 \includegraphics[width=.5\linewidth]{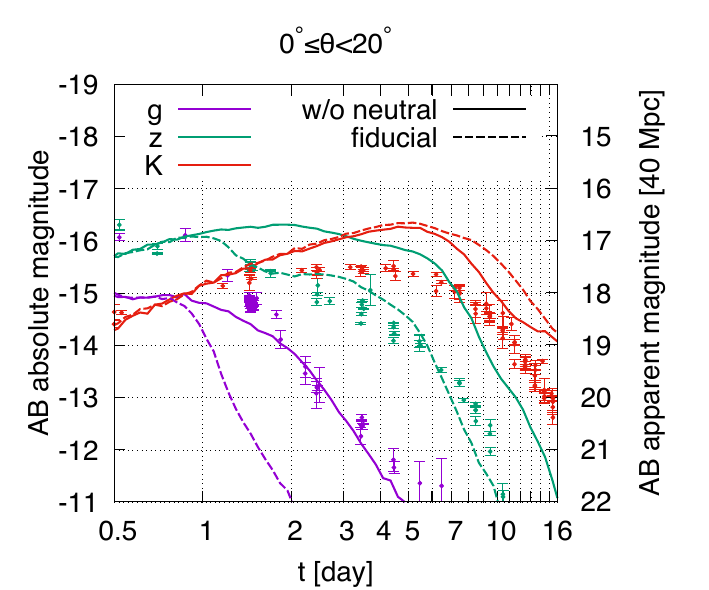}
 	 \caption{(Left panel) Spatial distribution of the average ionization state at $t=1.5\,{\rm day}$ (Right panel). Comparison of the {\it gzK}-band light curves observed from $0^\circ\le\theta\le20^\circ$ between the fiducial model (the dashed curves) and the model calculated by omitting the contribution of neutral atoms in the opacity calculation (the solid curves). The meaning of the data points is the same as in Figure~\ref{fig:mag_fid}.}
	 \label{fig:ion_fid}
\end{figure*}

The left panel of Figure~\ref{fig:ion_fid} shows the profile of the average ionization level at 1.5 days defined by $\sum_{n,m} m X_{n,m}$, where $X_{n,m}$ denotes the mass fraction of the $n$-th element in the $m$-th ionization state. The figure shows that the neutral atoms have already started being the dominant components in the polar region at 1.5 day due to relatively low heating rates and resulting low temperature (see Figure~\ref{fig:nuc_fid}). We find that neutral atoms become the dominant elements in the entire ejecta except the central region of which velocity is below $\approx 0.1$--$0.2\,c$ after the following few days. While most of the ejecta mass is concentrated in the central region and the energy source of the emission is dominated by the ejecta in such a region, the surrounding material with higher velocity can still contribute as an opacity source and modifies the spectra by reprocessing photons emitted from the center.

To examine the effect due to the possible non-LTE effect, we perform a radiative transfer simulation with the contribution of neutral atoms to the opacity being suppressed. For this setup, after the ionization population is determined by solving Saha's equation, we artificially change neutral atoms to the first ionized atoms and used such modified ionization population for the opacity calculation. The right panel of Figure~\ref{fig:ion_fid} compares the {\it gzK}-band light curves between the fiducial model and the model calculated by omitting the contribution of neutral atoms in the opacity calculation. We find that, while the peak magnitudes are not changed significantly, the brightness of the optical and near-infrared emission is sustained for a longer period for the case that neutral atoms are omitted.

\begin{figure}
 	 \includegraphics[width=\linewidth]{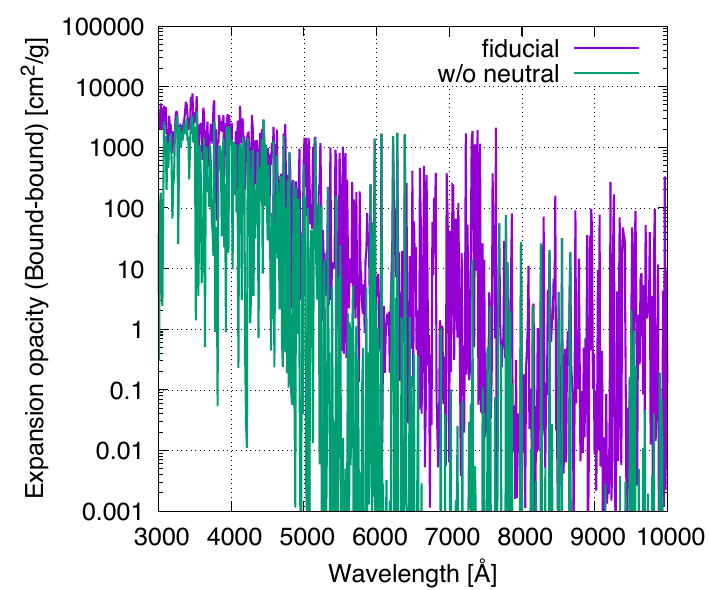}
 	 \caption{Comparison of the expansion opacity~\citep{1993ApJ...412..731E,Kasen:2006ce} of bound-bound transitions between the fiducial model (the blue curve) and the model calculated by omitting the contribution of neutral atoms (the green curve). The opacity is calculated by employing the element abundances in $(x/ct,z/ct)=(0.001,\,0.4)$ and by assuming $6\times 10^{-17}{\,\rm g/cm^3}$ and $2300\,$K, which correspond to the condition at $t=\,2$ days.}
	 \label{fig:opc_ionmod}
\end{figure}

The enhancement in the optical emission found in the right panel of Figure~\ref{fig:ion_fid} can be understood by the wavelength dependence of the opacity. Figure~\ref{fig:opc_ionmod} compares the opacity in $(x/ct,z/ct)=(0.001,\,0.4)$ at $t=\,2$ days between the fiducial model and the model calculated by omitting the contribution of neutral atoms. The ionized atoms are typically more transparent for photons with the long wavelengths than the neutral atoms~\citep{Gaigalas:2019ptx,Tanaka:2019iqp}. Indeed, Figure~\ref{fig:opc_ionmod} shows that, in the absence of neutral atoms, the ejecta are less opaque for photons with the wavelengths above 5000 \AA ($\approx$ the {\it g}-band).

We note that the enhancement of the emission found in the right panel of Figure~\ref{fig:ion_fid} could be overestimated since the neutral atoms are artificially omitted by hand. There is also a possibility that not only the population of the neutral atoms but also the atoms in the higher ionization states can be modified by the non-LTE effects. To examine what kind of ionization population can be actually realized, and to quantify the difference from the results assuming the thermal population, the implementation of non-LTE effects to the radiative transfer code is necessary. We leave this as the future work.

\section{Summary}\label{sec:S}
In this work, we studied the long-term evolution of the ejecta formed in a BNS merger that results in a long-lived remnant NS by performing a HD simulation with the outflow data of a NR simulation~\citep{Fujibayashi:2020dvr} as the initial condition. We found that the ejected material exhibits a mildly prolate shape, while the ejecta with relatively high-lanthanide mass fraction ($\gtrsim0.01$) show torus-like morphology. The increase of the ejecta temperature due to the interaction between ejecta components during the late-time hydrodynamics evolution had only a small effect to the nucleosynthesis, which supports the robustness of the resulting element abundances obtained in the NR simulations. 
On the other hand, we showed that a fraction of the material counted as ejecta falls back to the central region and fails to escape from the system due to the pressure from the preceding material. This indicates that feed-back effects of the fall-back material to the NR simulations might need to be considered and investigated to predict the late-phase evolution of the system accurately (see ~\citealt{Fernandez:2014bra,Fernandez:2016sbf} for black hole-NS mergers). 

We performed a radiative transfer simulation based on the ejecta profile in the homologously expanding phase obtained by our long-term HD simulation. We found that a large amount of total ejecta with low lanthanide fraction does not always result in the bright optical emission. Indeed, the optical emission was not as bright as in AT2017gfo despite the inferred large amount of total ejecta mass and low lanthanide fraction of the ejecta. We showed that preferential diffusion of photons toward the equatorial direction due to the prolate ejecta morphology, large opacity contribution of Zr, Y, and lanthanides, and low specific heating rate of the ejecta are the keys for this light curve property. This indicates that the progenitor of AT2017gfo is not likely to be a BNS merger that results in a long-lived remnant NS by which a strong polar outflow is necessarily driven. Our non-trivial findings increase the importance of the realistic ejecta modeling by employing the NR simulation data for the kilonova light curve prediction. 

Since~\cite{Fujibayashi:2020dvr} suggested that ejecta from BNS mergers that result in long-lived remnant NSs will share the common property, the resulting kilonova light curves from such systems might show the similar property to what we found in this work: kilonovae with relatively faint optical and bright infrared emission. The future observation of a kilonova with such features could be a good indicator for the formation of a long-lived remnant NS. 

We pointed out that the presence of a bright optical emission in the kilonova might be the indicator for the collapse of the merger remnant NS to a black hole in a short time scale ($\sim 0.1$\,s) after the onset of merger. Indeed, we showed that the suppression of the high velocity ejecta components in the polar region will enhance the optical emission. However, it is not clear that such an ejecta profile is indeed realized for the case that the remnant NS collapses to a black hole because we do not confirm the results of this type by NR simulations. We also note that there have been proposed alternative ways that possibly realize the bright optical emission~\cite[see e.g.,][]{Piro:2017ayh,Arcavi:2018mzm,Matsumoto:2018mra,Nativi:2020moj,Klion:2020efn}. Thus, further systematic investigation based on NR simulations is needed to obtain the deeper understanding. The nucleosynthesis and resultant radioactive heating rates can also depend on the adopted nuclear ingredients such as the mass model (HFB-21 of~\citealt{Goriely:2010bm} in this study; see~\citealt{Fujibayashi:2020dvr}), as pointed out by, e.g.,~\cite{Wu:2018mvg,Zhu:2020eyk}. Furthermore, we also pointed out that more detailed physical process in radiative transfer, such as non-LTE effect to the ionization population, may also modify the results. This indicates that more detailed microphysics will be needed for the accurate light curve prediction.

\acknowledgments We thank Smaranika Banerjee, Kenta Hotokezaka, Wataru Ishizaki, and Koutarou Kyutoku for valuable discussions. Numerical computation was performed on Cray XC40 at Yukawa Institute for Theoretical Physics, Kyoto University and Sakura cluster at Max Planck Institute for Gravitational Physics (Albert Einstein Institute). This work was supported by Grant-in-Aid for Scientific Research (JP16H02183, JP17H06361, JP15H02075, JP17H06363, JP18H05859, JP19H00694, JP20H00158) of JSPS and by a post-K computer project (Priority issue No. 9) of Japanese MEXT.

\bibliographystyle{apj}
\bibliography{ref}

\appendix

\section{Formulation}\label{app:form}
In this appendix, we describe the formulation of axisymmetric hydrodynamics equations in the spherical coordinates employed for the long-term evolution of ejecta. Throughout this appendix, the units of $c=1=G$ are employed where $G$ is the gravitational constant, unless otherwise mentioned.
\subsection{Basic equations}
The basic equations for the numerical hydrodynamics employed in this work are formulated in the framework of the 3+1 decomposition of the spacetime~\citep[see e.g.,][]{Shibata2015}. In the 3+1 form, the metric tensor $g_{\mu\nu}$ is decomposed as
\begin{align}
	ds^2=g_{\mu\nu}dx^\mu dx^\nu=-\alpha^2dt^2+\gamma_{ij}\left(dx^i+\beta^idt\right)\left(dx^j+\beta^jdt\right),
\end{align}
where $\mu$ and $\nu$ denote the spacetime indices, $i$ and $j$ denote the spatial indices, $\alpha$, $\beta^i$, and $\gamma_{ij}$ denote the lapse, shift, and spatial metric, respectively. We treat the material as a perfect fluid and the energy-momentum tensor is given by
\begin{align}
	T_{\mu\nu}=\rho hu_\mu u_\mu+Pg_{\mu\nu},
\end{align}
where $\rho$, $h$, $u^\mu$, and $P$ denote the rest-mass density, specific enthalpy, four velocity, and pressure, respectively. The equations of energy-momentum conservation and the continuity equation are given by
\begin{align}
	\gamma_{\nu i}\nabla_\mu T^{\mu\nu}&=\rho {\dot \epsilon} u_i\label{eq:eom}\\
	n_\nu \nabla_\mu T^{\mu\nu}&=\rho {\dot \epsilon} n_\nu u^\nu\label{eq:eoe}\\
	\nabla_\mu \left(\rho u^\mu\right)&=0,\label{eq:eoc}
\end{align}
with the covariant derivative, $\nabla_\mu$. Here, $n_\nu=-\alpha\nabla_\nu t$, $\gamma_{\mu\nu}=g_{\mu\nu}+n_\mu n_\nu$, and ${\dot \epsilon}$ is the specific heating rate of the radioactive heating. Equations~\eqref{eq:eom}, \eqref{eq:eoe} and \eqref{eq:eoc} are rewritten in the forms 
\begin{align}
	\partial_t S_i+\partial_k\left(S_i v^k+P\alpha\sqrt{\gamma} \delta^k_i \right)=-S_0\partial_i\alpha+S_k\partial_i\beta^k-\frac{1}{2}\alpha\sqrt{\gamma}S_{jk}\partial_i\gamma^{jk}+\frac{\alpha{\dot \epsilon}}{hw}S_i,
\end{align}
\begin{align}
	\partial_t S_0+\partial_k\left[S_0 v^k+P\sqrt{\gamma} \left(v^k+\beta^k\right) \right]=-\gamma^{ij}S_i\partial _j \alpha+\alpha\sqrt{\gamma}S_{ij}K^{ij}+\alpha\rho_*{\dot \epsilon},
\end{align}
\begin{align}
	\partial_t \rho_*+\partial_k\left(\rho_*v^k\right)=0,
\end{align}
respectively. Here, $K_{ij}$ denotes the extrinsic curvature, and the other variables which newly appear in the above equations are defined as follows:
\begin{align}
\sqrt{\gamma}&={\rm det}\left(\gamma_{ij}\right),\nonumber\\
\rho_*&=\rho w\sqrt{\gamma},\nonumber\\
w&=\alpha u^t,\nonumber\\
S_i&=\rho_* {\hat u}_i=\rho_* h u_i,\nonumber\\
S_0&=\rho_* {\hat e}=\rho_*\left(hw-\frac{P}{\rho w}\right),\nonumber\\
S_{ij}&=\rho hu_iu_j+P\gamma_{ij},\nonumber\\
v^i&=\frac{u^i}{u^t}.\label{eq:varidef}
\end{align}

\subsubsection{Basic equations in the spherical coordinates for the axisymmetric system}
In our calculation, the hydrodynamics equations are solved in the fixed background and imposing the axisymmetry. The basic equations in the spherical coordinates are written as
\begin{align}
	\partial_t \left(r^2{\rm sin}\theta {\tilde \rho}_*\right)+\partial_r\left(r^2 {\rm sin}\theta {\tilde \rho}_* v^{(r)}\right)+\partial_\theta\left(r {\rm sin}\theta {\tilde \rho}_* v^{(\theta)}\right)=0,
\end{align}
\begin{align}
	\partial_t \left(r^2{\rm sin}\theta {\tilde S}_{(r)}\right)&+\partial_r\left[r^2 {\rm sin}\theta\left( {\tilde S}_{(r)} v^{(r)}+P\alpha\sqrt{{\tilde \gamma}}\right)\right]+\partial_\theta\left(r {\rm sin}\theta {\tilde S}_{(r)} v^{(\theta)}\right)\nonumber\\
	&=r^2{\rm sin}\theta\left[-{\tilde S}_0\partial_r \alpha+{\tilde S}_{(i)} \partial_r \beta^{(i)}-\frac{1}{2}\alpha\sqrt{{\tilde \gamma}}{\tilde S}_{(i)(j)}\partial_r {\tilde \gamma}^{(i)(j)}\right.\nonumber\\
	&\left.+\frac{2}{r}\alpha\sqrt{{\tilde \gamma}} P +\frac{1}{r}\left({\tilde S}_{(\theta)}v^{(\theta)}+{\tilde S}_{(\varphi)}v^{(\varphi)}\right)+\frac{\alpha}{hw} {\tilde S}_{(r)}{\dot \epsilon} \right],
\end{align}

\begin{align}
	\partial_t \left(r^3{\rm sin}\theta {\tilde S}_{(\theta)}\right)&+\partial_r\left(r^3 {\rm sin}\theta {\tilde S}_{(\theta)} v^{(r)}\right)+\partial_\theta \left[r^2 {\rm sin}\theta\left( {\tilde S}_{(\theta)} v^{(\theta)}+P\alpha\sqrt{{\tilde \gamma}}\right)\right]\nonumber\\
	&=r^3 {\rm sin}\theta\left[-{\tilde S}_0\frac{1}{r}\partial_\theta \alpha+{\tilde S}_{(i)} \frac{1}{r}\partial_\theta \beta^{(i)}-\frac{1}{2}\alpha\sqrt{{\tilde \gamma}}{\tilde S}_{(i)(j)}\frac{1}{r}\partial_\theta {\tilde \gamma}^{(i)(j)}\right.\nonumber\\
	&\left.+\frac{1}{r}\alpha\sqrt{{\tilde \gamma}} P {\rm cot}\theta+\frac{1}{r}{\tilde S}_{(\varphi)}v^{(\varphi)}{\rm cot}\theta +\frac{\alpha}{hw} {\tilde S}_{(\theta)}{\dot \epsilon} \right],
\end{align}
\begin{align}
	\partial_t \left(r^3{\rm sin}^2\theta {\tilde S}_{(\varphi)}\right)&+\partial_r\left(r^3 {\rm sin}^2\theta {\tilde S}_{(\varphi)} v^{(r)}\right)+\partial_\theta\left(r^2 {\rm sin}^2\theta {\tilde S}_{(\varphi)} v^{(\theta)}\right)=r^3{\rm sin}^2\theta \frac{\alpha}{hw} {\tilde S}_{(\varphi)}{\dot \epsilon},
\end{align}
\begin{align}
	\partial_t\left( r^2{\rm sin}\theta {\tilde S}_0\right)&+\partial_r\left\{r^2{\rm sin}\theta\left[{\tilde S}_0v^{(r)}+P\sqrt{{\tilde \gamma}}\left(v^{(r)}+\beta^{(r)}\right)\right]\right\}\nonumber\\
	&+\partial_\theta\left\{r{\rm sin}\theta\left[{\tilde S}_0v^{(\theta)}+P\sqrt{{\tilde \gamma}}\left(v^{(\theta)}+\beta^{(\theta)}\right)\right]\right\}\nonumber\\
	&= r^2{\rm sin}\theta\left[-{\tilde \gamma}^{(i)(j)}{\tilde S}_{(i)}\partial_{(j)} \alpha+\alpha\sqrt{\tilde {\gamma}}{\tilde S}_{(i)(j)}K^{(i)(j)}+\alpha {\tilde \rho}_* {\dot \epsilon}\right].
\end{align}
Here, the definitions for the new variables are given as follows:
\begin{align}
{\tilde \gamma}_{(i)(j)}&=\Lambda_{(i)}^k\Lambda_{(j)}^l \gamma_{kl},\nonumber\\
{\tilde K}_{(i)(j)}&=\Lambda_{(i)}^k\Lambda_{(j)}^l K_{kl},\nonumber\\
\Lambda_{(r)}^r=1,~\Lambda_{(\theta)}^\theta=\frac{1}{r},&~\Lambda_{(\varphi)}^\varphi=\frac{1}{r{\rm sin}\theta},~\Lambda_{(i)}^j=0\,(i\ne j),\nonumber\\
\sqrt{{\tilde \gamma}}&=\frac{1}{r^2{\rm sin}\theta}\sqrt{\gamma},\nonumber\\
{\tilde \rho}_*&=\frac{1}{r^2{\rm sin}\theta} \rho_*,\nonumber\\
v^{(r)}=v^r,~v^{(\theta)}&=r v^\theta,~v^{(\varphi)}=r {\rm sin}\theta\, v^\varphi,\nonumber\\
\beta^{(r)}=\beta^r,~\beta^{(\theta)}&=r \beta^\theta,~\beta^{(\varphi)}=r {\rm sin}\theta\, \beta^\varphi,\nonumber\\
{\tilde S}_{(r)}=\frac{1}{r^2{\rm sin}\theta} S_r,~{\tilde S}_{(\theta)}&=\frac{1}{r^3{\rm sin}\theta} S_\theta,~{\tilde S}_{(\varphi)}=\frac{1}{r^3{\rm sin}^2\theta} S_\varphi,\nonumber\\
{\tilde S}_{(i)(j)}&=\Lambda_{(i)}^k\Lambda_{(j)}^l S_{kl}.
\end{align}
The indices without the parenthesis denote the tensor components with respect to the coordinate basis. We note that, in the flat spacetime, the indices with the parenthesis ($(i),\,i=r,\,\theta,\,\varphi$) denote the tensor components of the unit normal basis in the spherical coordinates. 

For the case of ${\dot \alpha}=0$, $\beta^{(i)}=0$, ${\tilde \gamma}_{(i)(j)}=\psi^4 \delta_{(i)(j)}$, and $K_{(i)(j)}=0$, which hold for the non-rotating black-hole spacetime in the isotropic coordinates, the equation of energy conservation can be simplified in the following form:
\begin{align}
	\partial_t\left( \alpha r^2{\rm sin}\theta {\tilde S}_0\right)&+\partial_r\left[\alpha r^2{\rm sin}\theta\left({\tilde S}_0v^{(r)}+P\sqrt{{\tilde \gamma}}v^{(r)}\right)\right]+\partial_\theta\left[\alpha r{\rm sin}\theta\left({\tilde S}_0v^{(\theta)}+P\sqrt{{\tilde \gamma}}v^{(\theta)}\right)\right]=\alpha^2 r^2{\rm sin}\theta {\tilde \rho}_* {\dot \epsilon}.
\end{align}
In this work, we solve the set of equations above by employing a Kurganov-Tadmor scheme~\citep{2000JCoPh.160..241K} with a piecewise parabolic reconstruction for the quantities of cell interfaces and the minmod-like filter introduced in~\cite{2000JCoPh.160..241K} for the flux-limitter.

\subsection{Test problems}\label{app:test}
\subsubsection{Point explosion (non-relativistic limit)}

\begin{figure*}
 	 \includegraphics[width=.5\linewidth]{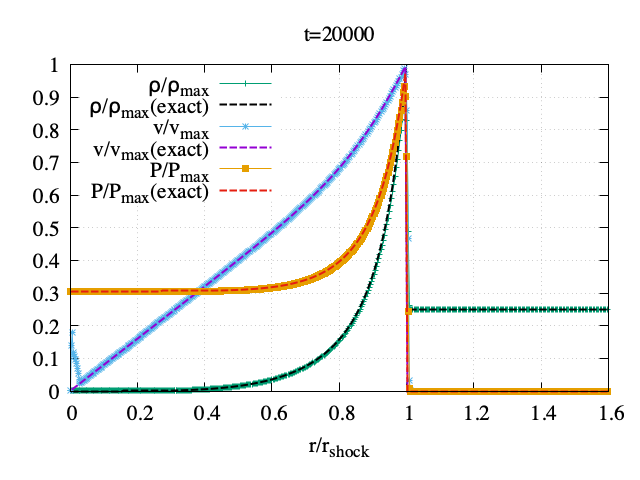}
 	 \includegraphics[width=.5\linewidth]{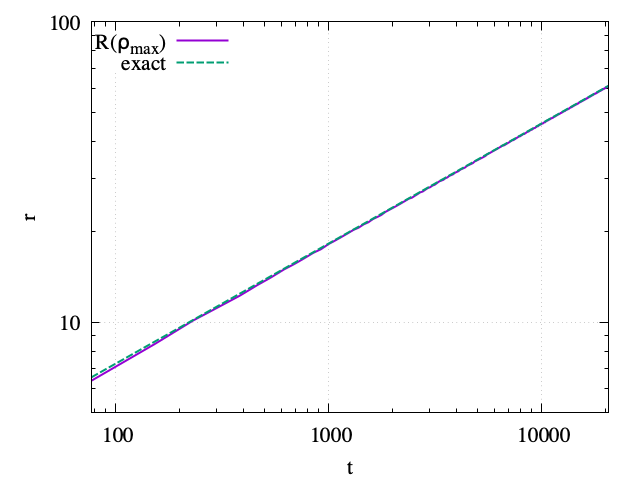}
 	 \caption{(Left panel) Radial distribution of the density, velocity, and pressure at $t=20000$ for the non-relativistic point explosion test problem (in arbitrary units). Those for the exact solution~\citep{von1941point,taylor1941british,sedov1946propagation} are plotted with the dashed curves. (Right panel) Time evolution of the location of the density maximum for the simulation (the blue solid) and the exact solution (the green dashed).}
	 \label{fig:test_sedov}
\end{figure*}

As the first test, we show that our code reproduces the analytic solution for the non-relativistic point explosion~\citep{von1941point,taylor1941british,sedov1946propagation}. We set the gas with uniform density of $\rho_0=1$, and inject $E_0=1$ of the internal energy uniformly in the region of $r\le 1$ (in arbitrary units). We employ the ideal gas EOS with the adiabatic index of $\Gamma=5/3$. The simulation is performed by employing the grid structure described in Equation~\eqref{fig:test_sedov} with $J=500$, $r_{\rm min}=0$, and $r_{\rm max}=100$. Figure~\ref{fig:test_sedov} shows the radial distribution of the density, velocity, and pressure at $t=20000$ as well as the location of the density maximum as a function of time with those obtained by the exact solution~\citep{von1941point,taylor1941british,sedov1946propagation}. We check that the L1 norm error between the conserved mass density of the simulation, $\rho_*$ (see Equation~\eqref{eq:varidef}), and the exact solution, $\rho_{\rm *,exact}$, which we define by
\begin{align}
	I:=\frac{\int \left|\rho_*-\rho_{\rm *,exact}\right|d^3x}{\int \rho_{\rm *,exact} d^3x},\label{eq:l1norm}
\end{align}
is smaller than $2\%$ and the shock location agrees with the exact solution within $\lesssim 0.5\%$ at $t=20000$.

\subsubsection{Homologously expanding ejecta}
As the second test problem, we inject the outflow profile of a homologously expanding ejecta to the inner boundary to examine whether our code can evolve the homologously expanding phase appropriately. The outflow which corresponds to the homologously expanding ejecta with the following density profile is employed:
\begin{align}
	\rho(v,\theta)=\left\{
	\begin{array}{cc}
	\rho_0 \left(\frac{v}{0.025\,c}\right)^{-3}\left\{0.01+\frac{0.99}{1+{\rm exp}\left[-20(\theta-\pi/4)\right]}\right\}&~(0.025\,c\le v\le 0.3\,c)\\
	\rho_0 \left(\frac{v}{0.025\,c}\right)^{6}\left\{0.01+\frac{0.99}{1+{\rm exp}\left[-20(\theta-\pi/4)\right]}\right\}&~(0.01\,c\le v\le 0.025\,c)\\
	0 &~({\rm else})
	\end{array}\right..
\end{align}

The total mass is set to be $\approx 7.4\times10^{-3} M_\odot$. For this examination, the ideal gas EOS with the adiabatic index of $\Gamma=4/3$ is employed, and we switch off the effect of gravity by setting the black-hole mass to be $0$. The same grid structure as in the fiducial HD simulation is employed for this test.

\begin{figure*}
 	 \includegraphics[width=.33\linewidth]{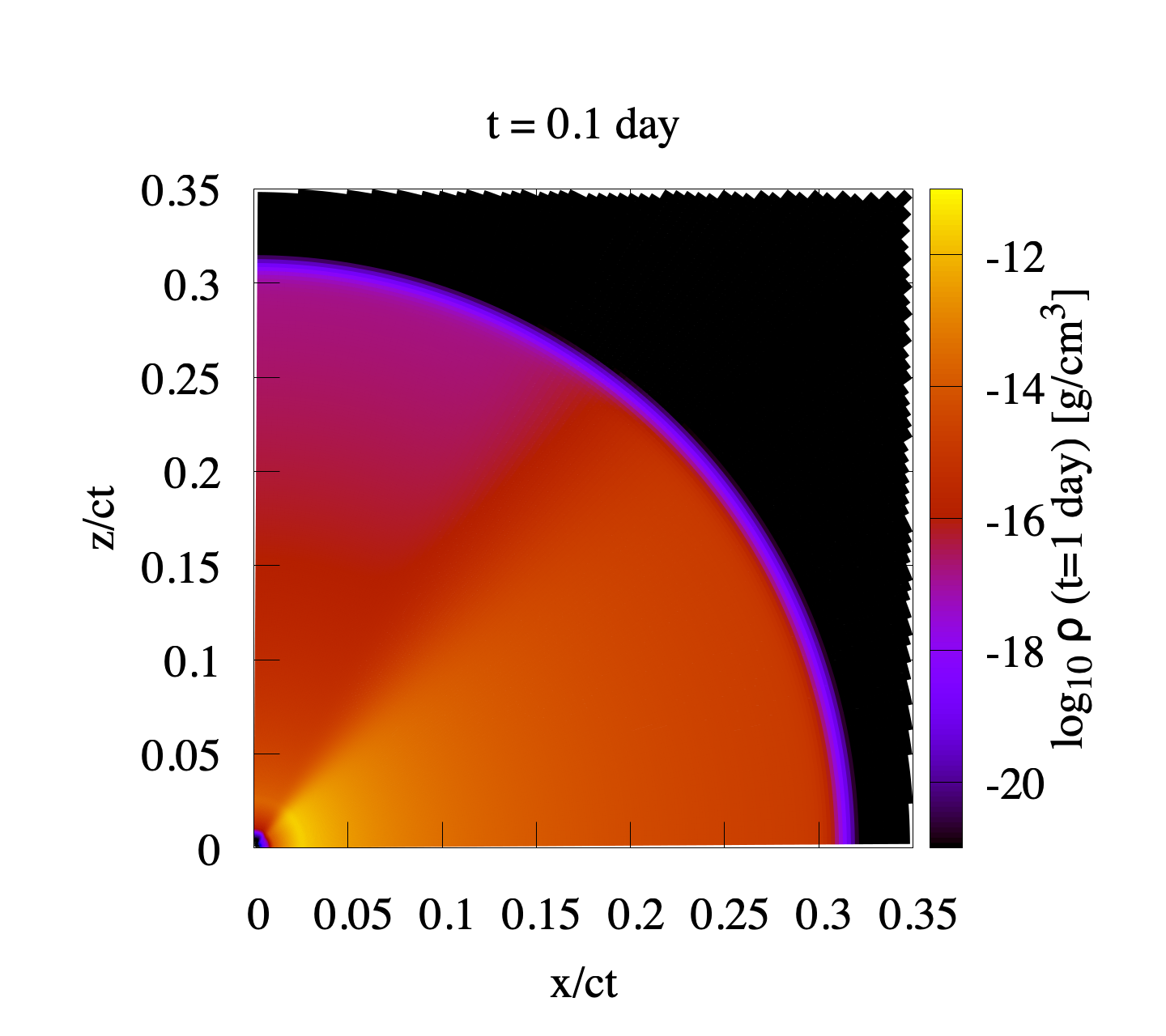}
 	 \includegraphics[width=.33\linewidth]{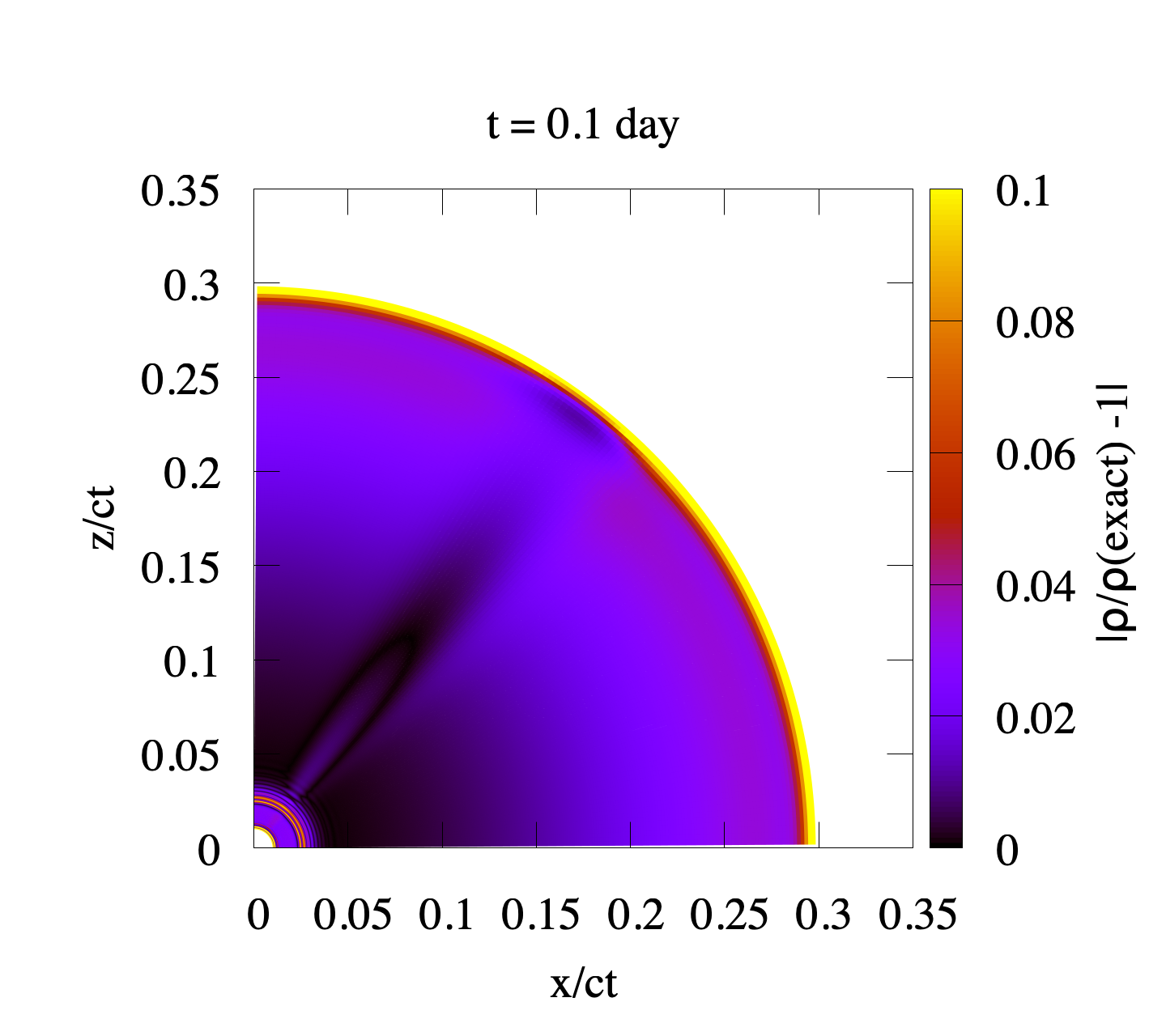}
 	 \includegraphics[width=.33\linewidth]{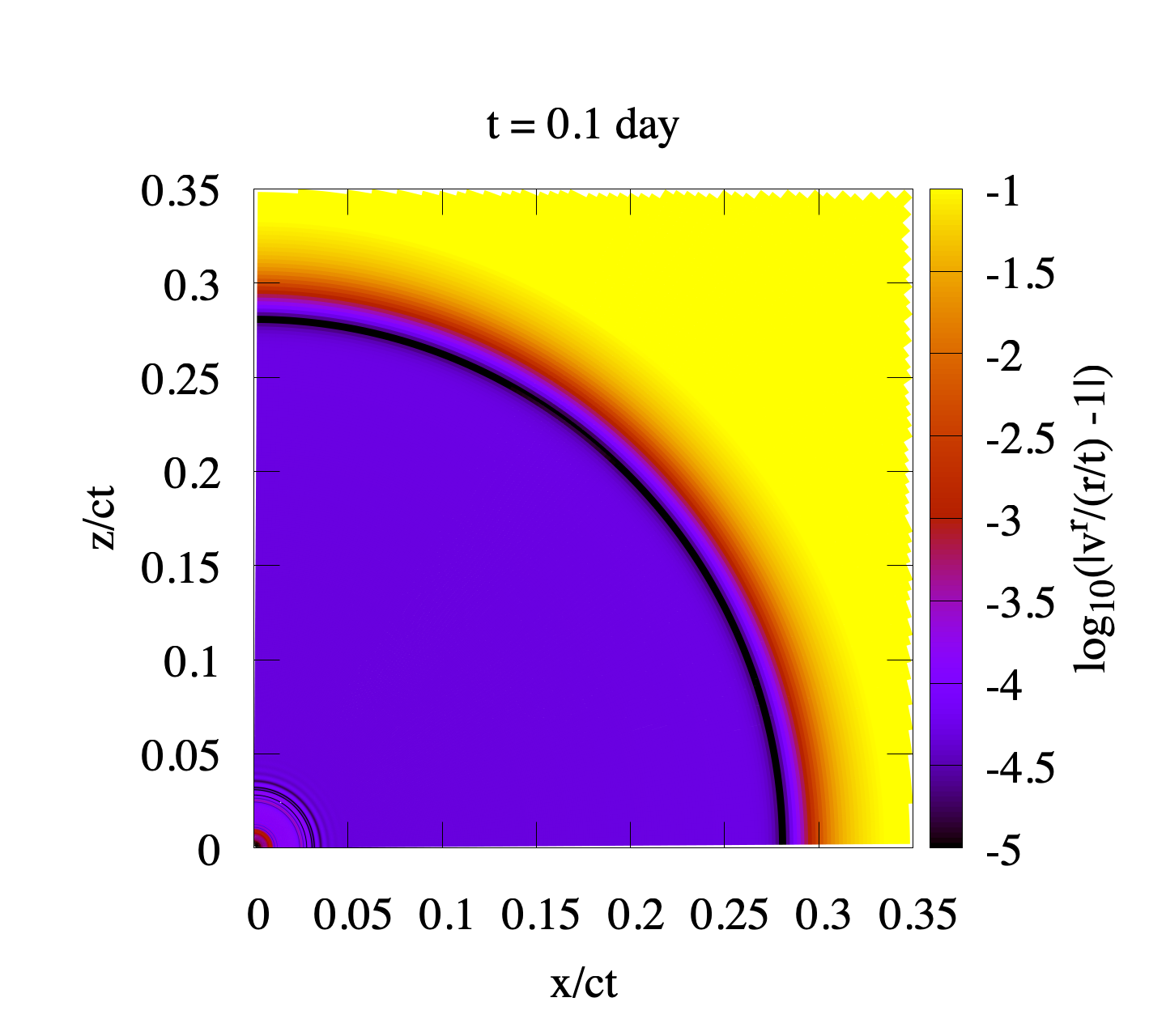}
 	 \caption{(Left panel) Rest-mass density profile of the homologously expanding ejecta obtained by the snapshot at $t=0.1$ days. (Middle panel) Relative difference of the rest-mass density profile from the exact solution at $t=0.1$ days. (Right panel) Relative difference of the radial velocity profile from that assuming the homologous expansion.}
	 \label{fig:test_homo}
\end{figure*}
Figure~\ref{fig:test_homo} shows the rest-mass density profile, relative difference of the rest-mass density profile from the exact solution, and relative difference of the radial velocity profile from that assuming the homologous expansion obtained by the snapshot at $t=0.1$ days. We confirm that the ejecta profile keeps the homologous feature for a sufficiently long period with reasonable accuracy by checking that the L1 norm error between the simulation result and the exact solution defined in Equation~\eqref{eq:l1norm} is $\approx 3\%$ and the deviation of the velocity from that for the homologous expansion is less than $0.3\%$ at $t=0.1$ days.

\subsubsection{Black hole torus}
Finally, to examine whether the effect of gravity is taken into account appropriately, we evolve the axisymmetric isentropic equilibrium state of a black hole-torus system. The initial profile is prepared in the same way as in~\cite{Fujibayashi:2020qda,Fujibayashi:2020jfr} with $n=1/7$ but for the fixed non-rotating black-hole background. The inner and outer edges of the torus are set to be $15\,M_{\rm BH}$ and $30\,M_{\rm BH}$, respectively, and the total mass of the torus is set to be $10^{-3}\,M_{\rm BH}$ with $M_{\rm BH}$ being the black hole mass. For this examination, the ideal gas EOS with the adiabatic index of $\Gamma=4/3$ is employed. The following grid structure is employed for the radial direction:
\begin{align}
	r_j=\left[\left(\sqrt{r_{\rm out}}-\sqrt{r_{\rm in}}\right)\frac{j-1}{J}+\sqrt{r_{\rm in}}\right]^2,\,j=1\cdots J+1\label{eq:grid2}
\end{align}
with $J=256$, $r_{\rm min}=10\,M_{\rm BH}$, and $r_{\rm max}=40\,M_{\rm BH}$. For $\theta$ direction, 128 grids uniformly covering the angle from $\theta=\pi/3$ to $\pi/2$ are employed. 
\begin{figure*}
 	 \includegraphics[width=.5\linewidth]{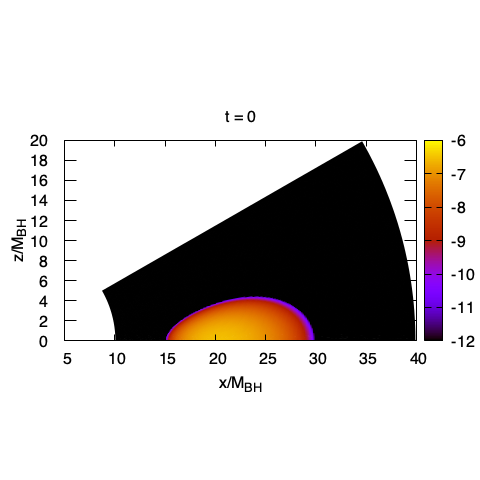}
 	 \includegraphics[width=.5\linewidth]{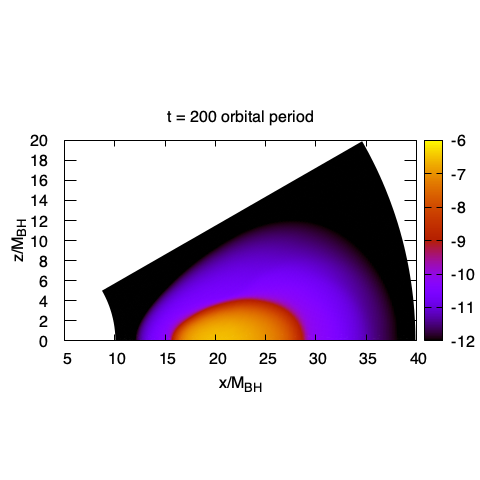}
 	 \caption{Rest-mass density profiles of a black hole-torus simulation at the initial time (the left panel) and at the time of the 200 orbital period at the rest-mass density maximum (the right panel).}
	 \label{fig:test_bhtorus}
\end{figure*}

Figure~\ref{fig:test_bhtorus} shows the rest-mass density profiles of a black hole-torus simulation at the initial time and at the time of the 200 orbital period at the rest-mass density maximum. Except for the small leak of the torus material into the atmosphere due to the numerical diffusion, our code indeed keeps the profile of the stationary solution for a long period. The L1 norm error of the conserved mass density between the initial profile and that at the time of the 200 orbital period is smaller than $\approx4\%$. We also check that the maximum rest-mass density always keeps the original value within $1\,\%$ during the simulation.

\section{Effect of the truncation in the outflow injection}\label{app:ext}

\begin{figure*}
 	 \includegraphics[width=.5\linewidth]{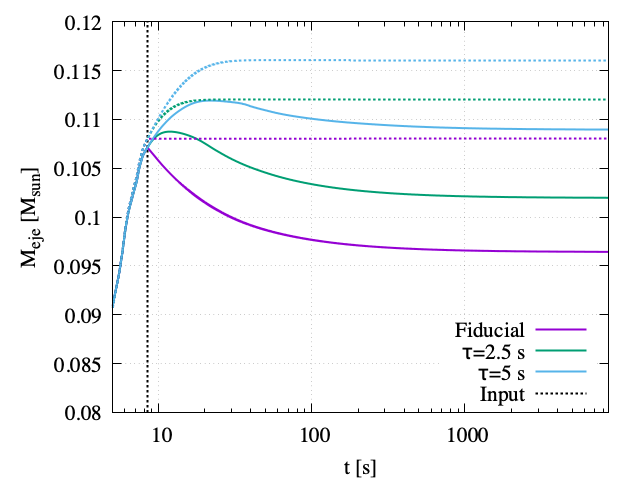}
 	 \includegraphics[width=.5\linewidth]{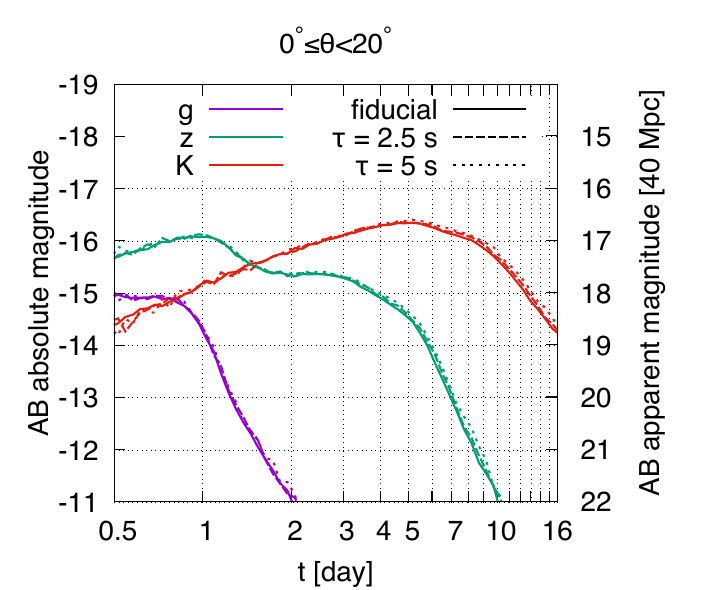}
 	 \caption{(Left panel) Total mass in the computational domain as a function of time for the models with the extended outflow injection. The purple, green, and blue curves denote the result for the models in which the outflow injection is truncated after the time at the end of the outflow data, extended by the time scale of $\tau=2.5$\,s, and extended by $\tau=5$\,s, respectively. The dotted curves denote the total input mass determined by the outflow data. The black dotted vertical line denotes the time at which the material injection from the inner boundary is terminated. (Right panel) Comparison of the {\it gzK}-band light curves between the fiducial ejecta model (the solid curves) and the ejecta model in which the outflow injection is extended by the time scale of $\tau=2.5$\,s (the dashed curves) and by $\tau=5$\,s (the dotted curves).}
	 \label{fig:meje_ext}
\end{figure*}

To investigate how the result depends on the way of truncating the ejecta injection, we perform the HD simulations in which the outflow injection from the inner boundary is continued even after the outflow data run out. Here, the ejecta profile at the end of the outflow data but with the density value suppressed by ${\rm exp}\left[-(t-t_{\rm end})/\tau\right]$ is employed for extended outflow injection, where $t_{\rm end}$ and $\tau$ are the end time of the outflow data and the time scale of extension, respectively. This prescription is justified by the fact that ejecta in the late phase ($t>5$ s) indicate approximately the same property~\citep{Fujibayashi:2020dvr}. Note that the fiducial setup can be regarded as the case of $\tau\rightarrow0$. 

Figure~\ref{fig:meje_ext} shows the late-time evolution of the total mass in the computational domain for the fiducial setup and for the cases with $\tau=2.5$\,s and $\tau=5$\,s. The total outflow mass computed from the mass flux on the inner boundary is $0.108\,M_\odot$, $0.112$ $M_\odot$ and $0.116$ $M_\odot$, respectively. Note that the total input mass for $\tau=5$\,s agrees approximately with the ejecta mass that measured in the NR simulation including the bulk component which still not yet escaped from the extraction radius at the end of the NR simulation ($\approx0.114\,M_\odot$, see Table 3 in~\citealt{Fujibayashi:2020dvr}). Since the ejecta formation seemed to be terminated at the end of the NR simulation (see the dashed curve in the top left panel of Figure 5 in~\citealt{Fujibayashi:2020dvr}), we expect that the case of $\tau=5$\,s mimics the outflow data obtained by the NR simulation in which the calculation is continued until the entire ejected material escapes from the extraction radius.

The mass in the computational domain only reaches 0.107 $M_\odot$, 0.108 $M_\odot$, and 0.112 $M_\odot$ at the peaks for the fiducial model and for the cases with $\tau=2.5$\,s and $\tau=5$\,s, respectively, and they are smaller than the total mass computed from the mass flux on the inner boundary.
This indicates that a fraction of the material falls back from $r=r_{\rm in}$ even though the outflow injection is continued. The mass in the computational domain decreases after reaching the peak, and it converges to 0.096 $M_\odot$, 0.102 $M_\odot$, and 0.109 $M_\odot$, respectively. A smaller fraction of the material falls back for the case with a larger value of $\tau$. This indicates that a larger amount of material in the bound orbits is accelerated for the model with the outflow injection extended for a longer time scale. Nevertheless, the mass of the ejecta which reaches the homologously expanding phase only varies by $\approx 10\%$. Moreover, as we show in the right panel of Figure~\ref{fig:meje_ext}, we find that the resulting kilonova light curves particularly for $t\leq6$ days do not strongly depend on how long the outflow data are extended. This is because the material additionally injected by the outflow extension typically has a small value of velocity ($\lesssim 0.05\,c$), and the diffusion time scale of photons emitted from it is long ($\gtrsim 6$ days).

\section{Effect of heating in the hydrodynamics evolution}\label{app:heat}

\begin{figure*}
 	 \includegraphics[width=.5\linewidth]{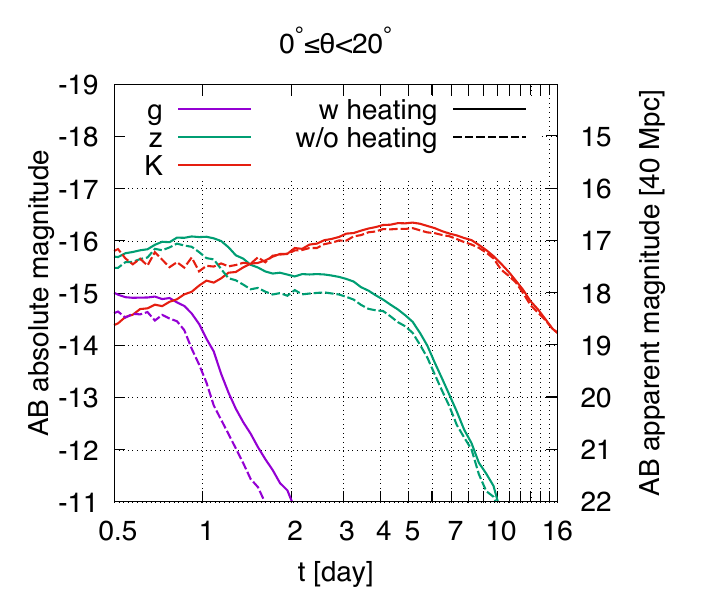}
 	 \includegraphics[width=.5\linewidth]{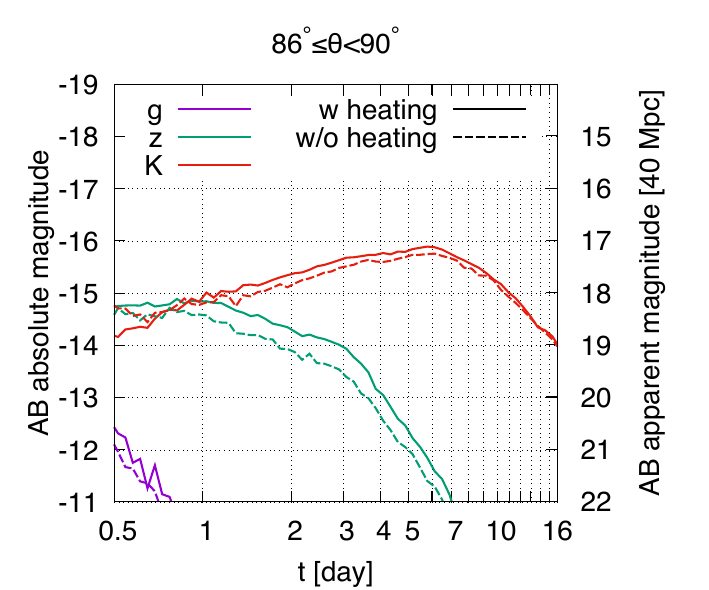}
 	 \caption{Comparison of the {\it gzK}-band light curves between the fiducial ejecta model (the solid curves) and the ejecta model calculated by omitting the heating source term for the HD simulation (the dashed curves). The light curves observed from $0^\circ\le\theta\le20^\circ$ (the left panel) and $86^\circ\le\theta\le90^\circ$ (the right panel) are shown.}
	 \label{fig:magcomp_heat}
\end{figure*}

Figure~\ref{fig:magcomp_heat} compares the {\it gzK}-band light curves between the fiducial ejecta model and the ejecta model calculated by omitting the heating source term in the HD simulation. The emission for the ejecta model calculated by omitting the heating source term is slightly fainter than that for the fiducial model. This is due to the difference in the internal energy initially set in the radiative transfer simulations taking over the results of the HD simulations. Indeed, we confirm this by checking that the light curves for the ejecta models obtained by the HD simulation without the heating source term agree approximately with those for the fiducial model if the same amount of internal energy is initially imposed.

\section{Effect of underestimating the internal energy}\label{app:intu14}
\begin{figure*}
 	 \includegraphics[width=.5\linewidth]{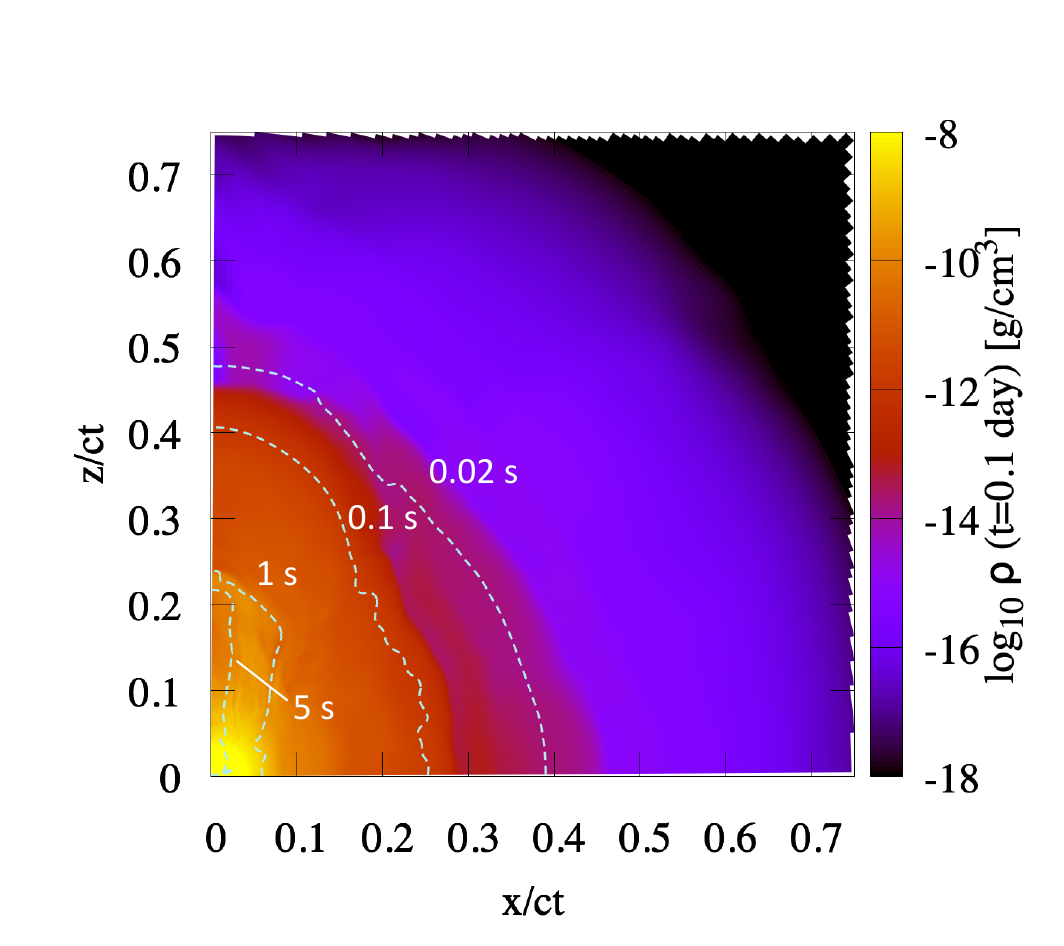}
	 \includegraphics[width=.5\linewidth]{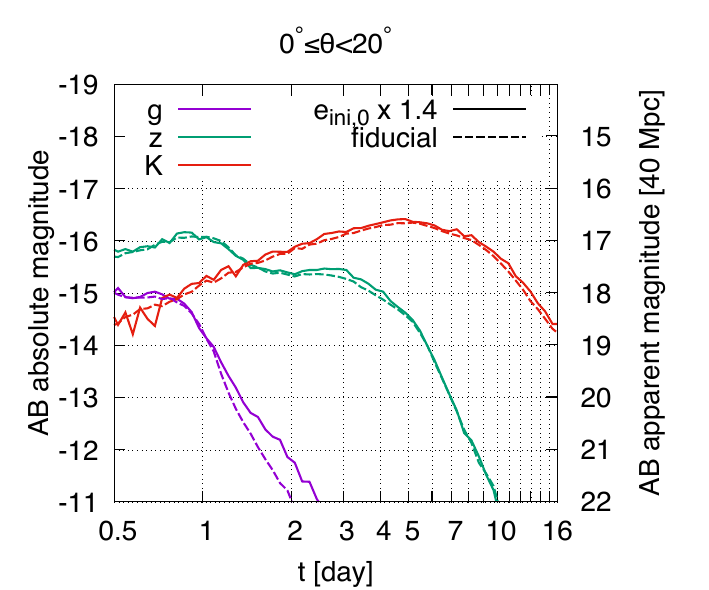}
 	 \caption{(Left panel) Rest-mass density profile of the ejecta at $t\approx 0.1\,{\rm days}$ for the HD simulation in which the injected internal energy is enhanced by $40\%$. The dashed curves denote the ejecta which escape from the extraction radius of the NR simulation at $t=0.02$, $0.1$, $1$, and $5\,{\rm s}$. (Right panel) Comparison of the {\it gzK}-band light curves observed from $0^\circ\le\theta\le20^\circ$ between the fiducial ejecta model (the dashed curves) and the ejecta model in which the injected internal energy is enhanced by $40\%$ (the solid curves).}
	 \label{fig:magcomp_u14}
\end{figure*}

As is discussed in Section~\ref{sec:hydro:heat}, the radioactive heating after the temperature drops below $\approx6$--$7$ GK is not taken into account in the NR simulation and the internal energy is  underestimated typically by $\approx40\%$ at the time which the material reaches the extraction radius of the NR simulation. To check how this affects the results, we performed the HD simulation and the radiative transfer simulation with the initial internal energy of the HD simulation (at the time which the material reached $r=r_{\rm in}$) increased by $40\%$.

The left panel of Figure~\ref{fig:magcomp_u14} shows the rest-mass density profile of the ejecta at $t\approx 0.1\,{\rm days}$ for the HD simulation in which the injected internal energy is enhanced by $40\%$. Comparing with Figure~\ref{fig:prof_dens}, the ejecta is slightly extended in the velocity space due to the acceleration induced by the increase in the internal energy. Nevertheless, we find that the acceleration of the ejecta velocity due to increase in the internal energy is only within $\approx 5\%$.

The right panel of Figure~\ref{fig:magcomp_u14} compares the {\it gzK}-band light curves observed from $0^\circ\le\theta\le20^\circ$ between the fiducial ejecta model and the ejecta model in which the injected internal energy is enhanced by $40\%$. We find that the lightcurves between these two models approximately agree with each other. Note that the slight enhancement in the {\it g}-band magnitude for $t\ge 1\,{\rm day}$ is due to the slightly extended density profile of the model with the enhanced internal energy injection.

\section{Homologously expanding phase}\label{app:homo}

\begin{figure*}
\center
 	 \includegraphics[width=.5\linewidth]{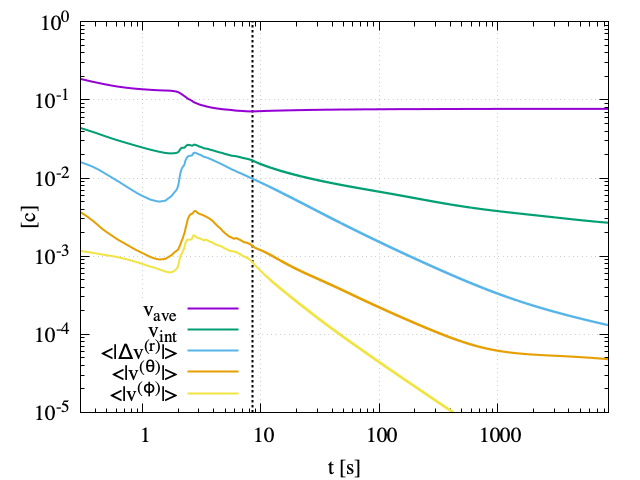}
 	 \caption{Comparison of $v_{\rm ave}$, $v_{\rm int}$, $\left<\left|\Delta v^{(r)}\right|\right>$, $\left<\left|v^{(\theta)}\right|\right>$, and $\left<\left|v^{(\varphi)}\right|\right>$ (see the main text for their definitions). The black dotted vertical line denotes the time at which the material injection from the inner boundary is truncated.}
	 \label{fig:v_time}
\end{figure*}

To examine at which time the ejecta reach the homologously expanding phase, we define
\begin{align}
v_{\rm ave}&=\sqrt{2 \frac{\int e_{\rm kin}d^3x}{\int \rho_* d^3x}},\\
v_{\rm int}&=\sqrt{2 \frac{\int e_{\rm int}d^3x}{\int \rho_* d^3x}},\\
\left<\left|\Delta v^{(r)}\right|\right>&=\frac{\int \rho_* w\left|v^{(r)}-r/t\right| d^3x}{\int \rho_* d^3x},\\
\left<\left|v^{(\theta)}\right|\right>&=\frac{\int \rho_* w \left|v^{(\theta)}\right|d^3x}{\int \rho_* d^3x},\\
\left< \left|v^{(\varphi)}\right|\right>&=\frac{\int \rho_* w \left|v^{(\varphi)}\right|d^3x}{\int \rho_* d^3x},
\end{align}
and compare the time evolution of these quantities in Figure~\ref{fig:v_time}. Here, $e_{\rm kin}=\rho_*\left(w-1\right)$ and $e_{\rm int}=\rho_*\left({\hat e}-w\right)$ denote the kinetic and internal energy density, respectively, and $v_{\rm ave}$, $v_{\rm int}$, $\left<\left|\Delta v^{(r)}\right|\right>$, $\left<\left|v^{(\theta)}\right|\right>$, and $\left<\left|v^{(\varphi)}\right|\right>$ denote the rms velocity calculated from the total kinetic energy, characteristic velocity calculated from the total internal energy, mass averaged deviation of the radial velocity from that assuming the homologous expansion, mass averaged absolute value of the latitudinal velocity, and mass averaged absolute value of the longitudinal velocity of ejecta, respectively. In the homologously expanding phase, we expect that $v_{\rm int}$, $\left<\left|\Delta v^{(r)}\right|\right>$, $\left<\left|v^{(\theta)}\right|\right>$, and $\left<\left|v^{(\varphi)}\right|\right>$ are smaller than $v_{\rm ave}$. Indeed, Figure~\ref{fig:v_time} shows that such a condition is satisfied for $t \gtrsim 1000$ s since $v_{\rm ave}$ is larger than other values by more than an order of magnitude.

\begin{figure*}
 	 \includegraphics[width=.5\linewidth]{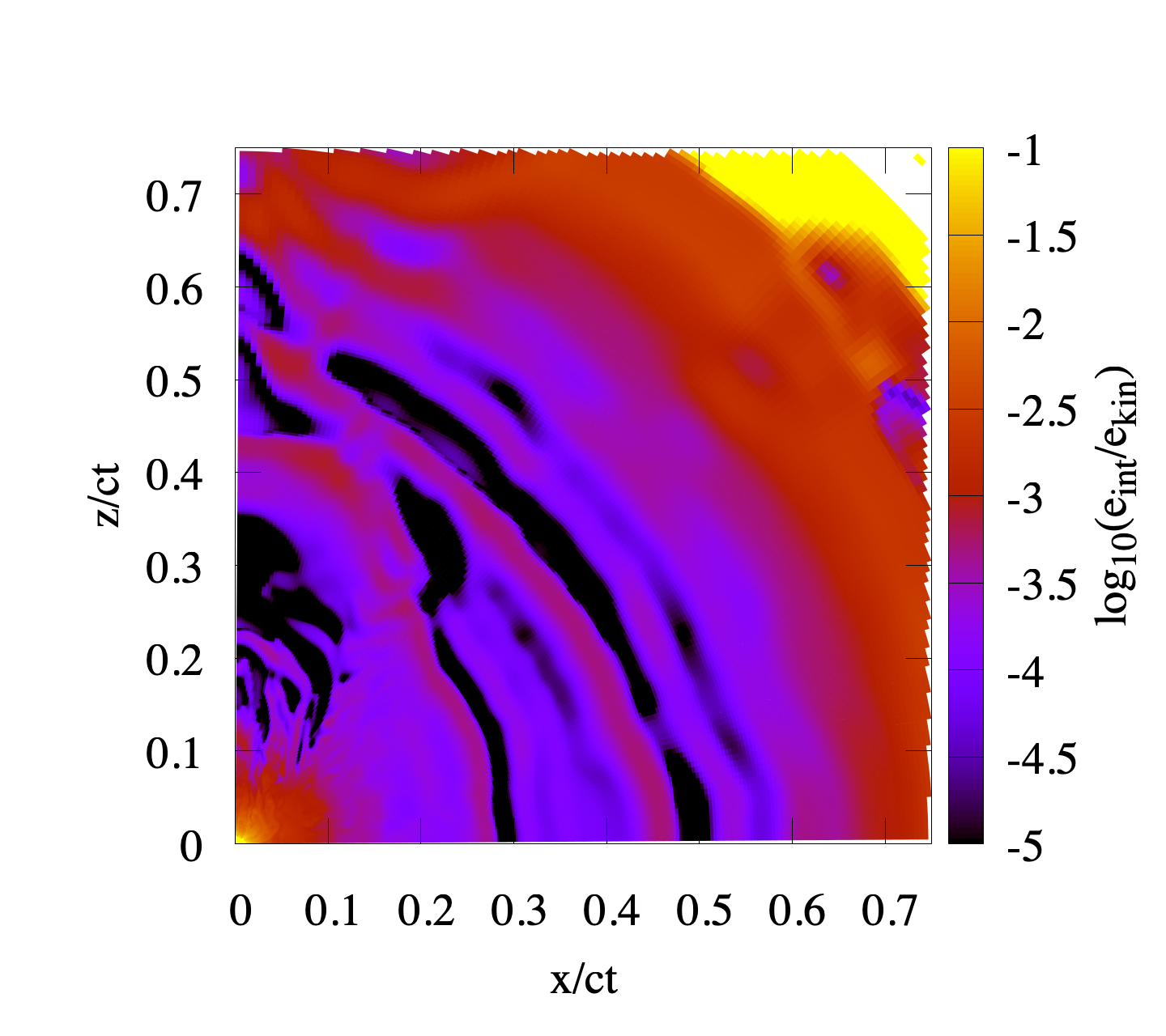} 
 	 \includegraphics[width=.5\linewidth]{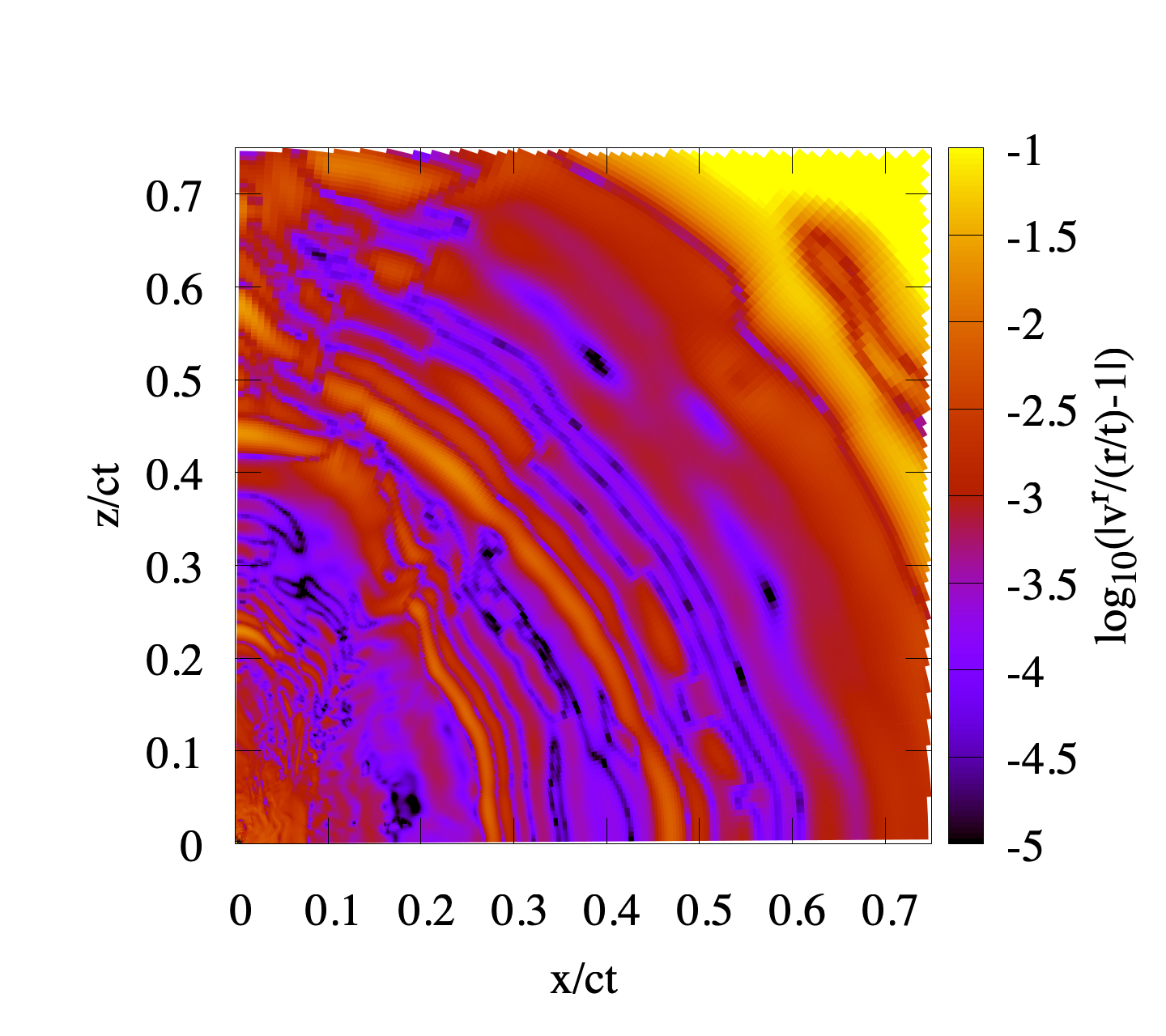}
 	 \caption{(Left panel) Ratio of the internal energy density to the kinetic energy density at $t\approx 0.1\,{\rm days}$. (Right panel) Relative difference of the radial velocity from that in the homologous phase at $t\approx 0.1\,{\rm days}$.}
	 \label{fig:homocheck_fid}
\end{figure*}
Figure~\ref{fig:homocheck_fid} shows the 2D profiles for the ratio of the internal energy density to the kinetic energy density (the left panel) and the relative difference of the radial velocity from that in the homologous phase (the right panel) at $t\approx 0.1\,{\rm days}$. The internal energy density is much smaller than the kinetic energy density by more than orders of magnitude, and the deviation of the radial velocity distribution from that assuming the homologous expansion is also smaller than $\approx 1\%$ for the entire ejecta. Thus, we can safely consider that the ejecta are in the homologously expanding phase at $t=0.1$ days.

\section{Higher viscosity model}\label{app:HV}
\begin{figure*}
 	 \includegraphics[width=.5\linewidth]{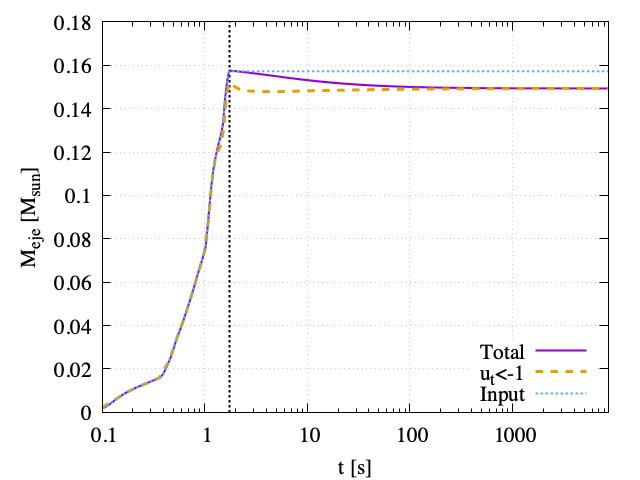}
 	 \includegraphics[width=.5\linewidth]{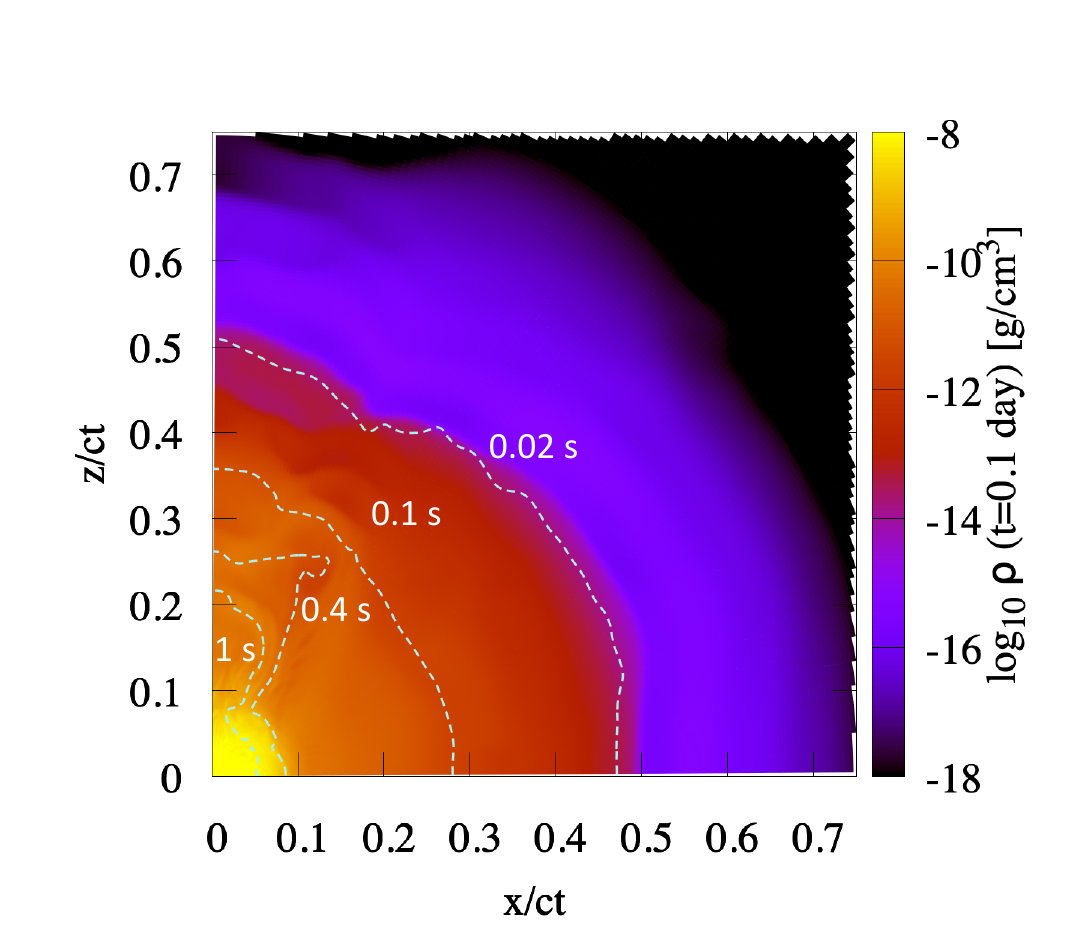}
 	 \caption{(Left panel) The same as Figure~\ref{fig:meje_time} but for the ejecta model employing the outflow data of the NR simulation with a large viscous parameter ($\alpha=0.10$). The black dotted vertical line denotes the time at which the material injection from the inner boundary is terminated. (Right panel) The same as the top panel of Figure~\ref{fig:prof_dens} but for the ejecta model employing the outflow data of the NR simulation with a large viscous parameter. The dashed curves denote the ejecta which escape from the extraction radius of the NR simulation at $t=0.02$, $0.1$, $0.4$, and $1\,{\rm s}$.}
	 \label{fig:meje_v10}
\end{figure*}
The left panel in Figure~\ref{fig:meje_v10} shows the time evolution of the total mass in the computational domain for the model employing the outflow data of the NR simulation with a large viscous parameter (``DD2-125M-h" in~\citealt{Fujibayashi:2020dvr}). The total input mass calculated from the mass flux of the outflow data is also shown in the figure. For the higher viscosity model, the outflow data run out at $t\approx1.7$ s. We note that $\approx0.05\,M_\odot$ of ejecta still remained inside the extraction radius of the NR simulation (=8000 km) at the end of the NR simulation. 

As is the case for the fiducial model, the total mass in the computational domain agrees with the total input mass calculated from the mass flux of the outflow data until the simulation time reaches the end of the outflow data. Two distinct mass ejection phases are seen: One found in the early phase ($t_{\rm in}\lesssim\,0.4$\,s) and the other found in the late phase ($t_{\rm in}\gtrsim\,0.4$\,s). Due to the larger viscous parameter ($\alpha_{\rm vis}=0.10$), the mass ejection occurs in a shorter time scale than for the fiducial model ($\alpha_{\rm vis}=0.04$). After the simulation time reaches the end of the outflow data, the total ejecta mass in the computational domain turns to decrease, and it converges to $\approx0.15\,M_\odot$ for $t\geq100$ s due to the vanishing pressure support from the inner boundary. The mass of the material in the unbound trajectory reaches the peak at the time when the outflow data run out, and slightly decreases subsequently. This subsequent decrease indicates the deceleration of the material due to the pressure from the preceding ejecta.

The right panel in Figure~\ref{fig:meje_v10} shows the rest-mass density profile of the ejecta obtained by the HD simulation at $t\approx 0.1\,{\rm days}$. Due to the larger viscous parameter, the ejecta are more extended to higher velocity than the fiducial model, and the early-time ejecta component ($t_{\rm in}\lesssim\,0.4$\,s) exhibits a mildly oblate shape. The late-time ejecta component ($t_{\rm in}\gtrsim\,0.4$\,s) shows approximately a prolate morphology as is the case for the fiducial model.

\end{document}